\documentclass[9pt,twocolumn,twoside]{pnas-new}

\templatetype{pnasresearcharticle} 
\usepackage{bm}
\begin{document}

\title{Sequence Complexity and Monomer Rigidity Control the Morphologies and Aging Dynamics of Protein Aggregates}
\author[a]{Ryota Takaki}
\author[d,e,1]{D. Thirumalai}

\affil[a]{Max Planck Institute for the Physics of Complex Systems, Nöthnitzer Str.38, 01187 Dresden, Germany}
\affil[b]{Department of Chemistry, The University of Texas at Austin, Austin, TX,78712}
\affil[c]{Department of Physics, The University of Texas at Austin, Austin, TX,78712}

\leadauthor{Takaki}

\significancestatement{Protein aggregates exhibit diverse morphology, exemplified by amyloid fibrils, gel-like structures, and liquid-like condensates. Differences in the  morphologies  in identical proteins play important functional roles in several diseases. Simulations using a minimal model show that such structures are encoded in the sequence complexity and bending rigidity of the monomers. The low-complexity flexible sequences form liquid droplets, whose relaxation dynamics are ergodic. In contrast, rigid low and high-complexity sequences, which form ordered nematic fibril-like structures and amorphous aggregates, exhibit heterogenous, non-ergodic dynamics.  The relaxation times under these conditions increase as the waiting time increases, which is a signature of aging. The implications of our findings for aging in intrinsically disordered proteins and repeat RNA sequences are outlined.}

\authorcontributions{RT and DT designed research, performed reseach, analyzed data, and wrote the paper}
\authordeclaration{The authors declare no competing interests.}
\correspondingauthor{\textsuperscript{1}To whom correspondence should be addressed. E-mail:dave.thirumalai@gmail.com}

\keywords{Biological Condensates $|$ Polymer Glass $|$ Slow Dynamics $|$ Ergodicity Breaking $|$ Dynamic Heterogeneity $|$ Trap Model $|$}

\begin{abstract}
Understanding the biophysical basis of protein aggregation is important in biology because of the potential link to several misfolding diseases. Although experiments have shown that  protein aggregates adopt a variety of morphologies, the dynamics of their formation are less well characterized. Here, we introduce a minimal model to explore the dependence of the aggregation dynamics on the structural and sequence features of the monomers. Using simulations we demonstrate that sequence complexity (codified in terms of word entropy) and monomer rigidity profoundly influence the dynamics and morphology of the aggregates. Flexible monomers with low sequence complexity  (corresponding to repeat sequences) form liquid-like droplets that exhibit ergodic behavior. Strikingly, these aggregates abruptly transition to more ordered structures, reminiscent of amyloid fibrils, when the monomer rigidity is increased. In contrast, aggregates resulting from monomers with high sequence complexity are amorphous and display non-ergodic glassy dynamics. 
The heterogeneous dynamics of the low and high-complexity sequences follow stretched exponential kinetics, which is one of the characteristics of glassy dynamics. Importantly, at non-zero values of the bending rigidities, the aggregates age with the relaxation times that increase with the waiting time. Informed by these findings, we provide insights into aging dynamics in protein condensates and contrast the behavior with the dynamics expected in RNA repeat sequences. Our findings underscore the influence of the monomer characteristics in shaping the morphology and dynamics of protein aggregates, thus providing a foundation for deciphering the general rules governing the behavior of protein condensates.
\end{abstract}

\dates{This manuscript was compiled on \today}
\doi{\url{www.pnas.org/cgi/doi/10.1073/pnas.XXXXXXXXXX}}

\maketitle
\thispagestyle{firststyle}
\ifthenelse{\boolean{shortarticle}}{\ifthenelse{\boolean{singlecolumn}}{\abscontentformatted}{\abscontent}}{}


\dropcap{P}hase behavior of polymer solution, which has become prominent in recent years owing to its relevance in biology,  is a classical problem in chemistry and physics~\cite{flory1953principles,de1979scaling,lifshitz1978some}.   The mean-field Flory-Huggins (FH) theory captures many aspects of phase separation of polymer-solvent mixtures~\cite{flory1942thermodynamics,huggins1942viscosity}, and has been widely applied to explain a large body of experimental data.  In particular, various aspects of polymer solution including gelation and the ability of polymer aggregates to form ordered phases have been theoretically studied~\cite{de1993physics, flory1941molecular,stockmayer1943theory,RubinsteinBook}. 
The FH theory, which has been extended to account for polyelectrolyte effects to describe complex coacervation~\cite{Overbeek57JCellComparativePhys}, also explains condensate formation in Intrinsically Disordered Proteins (IDPs) \cite{Brady17PNAS,Mccarty19JPCL}.

Although it has been known for a long time that proteins and RNA undergo phase separation~\cite{Overbeek57JCellComparativePhys,Eisenberg67JMB}, it is only after the discovery of the formation of membraneless Germline P granules~\cite{brangwynne2009germline}, that the significance of this phenomenon in biology has been adequately appreciated~\cite{alberti2019considerations}. These remarkable findings pose new challenges that are not nominally encountered in synthetic polymers. The most obvious is the precise connection between sequence and the macroscopic phase behavior, which is an active field of study~\cite{schuster2020identifying,lin2017random,regy2021improved}.  
Phase separation predominantly occurs in the low complexity domains of IDPs~\cite{martin2018relationship} or proteins with intrinsically disordered regions. 
For example, the low complexity domain of RNA binding proteins, Fused in Sarcoma (FUS) and hnRNPA1, not only results in condensate formation but also form ordered fibrils flanked by disordered brush-like coats~\cite{murray2017structure,lee2020molecular,molliex2015phase}.  Furthermore, protein condensates, which are initially liquid-like could acquire solid-like characteristics as they age, which is manifested by an increase in the relaxation time as the waiting time of the experiment increases~\cite{jawerth2020protein}. These observations suggest that the diagram of states as well as the dynamics of IDPs are complex and are determined by the various environmental factors (for example, pH change can induce a transition between fibrils and amorphous
aggregates~\cite{vetri2007amyloid,nishikawa2017two,yoshimura2012distinguishing}).   


It is likely that the formation of  
 liquid condensates, ordered fibrils and amorphous aggregates are controlled by both environment and monomer sequence. Indeed, several studies have addressed the relation between the protein sequences and resulting phase  behavior~\cite{dignon2018relation,harmon2017intrinsically,zeng2020connecting,rumyantsev2019controlling,lin2017phase,sawle2015theoretical,zhang2021decoding,tang2021prediction}. However, exactly solvable lattice models have shown that sequence gazing alone is insufficient to predict aggregation propensities~\cite{Li10PRL}. 
In this study, we pose two questions: What are the characteristic features of the monomers that lead to various forms of biomolecular aggregates, such as liquid condensates, amyloid fibrils, and amorphous solids?  Is there a connection between sequence complexity and the monomer energy landscape that could predict morphologies and dynamics (in particular aging) of aggregates? To answer these questions, we introduce a minimal model, the likes of which have been used previously to provide insights into several problems in protein aggregation~\cite{dima2002exploring,Li10PRL,Aggeli01PNAS}, and more recently in condensate formation~\cite{An24SciAdv,zhang2021decoding,Ronceray22PRL,Pyo23PNAS,Jacobs21PRL,Ranganathan20eLife}. 
Simulations using the minimal model show that variations in sequence complexity and rigidity of the monomers drastically alter the morphologies and the aging dynamics of aggregates. The low complexity flexibile monomers, like repeat RNA sequences,   form liquid droplets that exhibit ergodic behavior with no signs of aging. When the monomer binding rigidity increases the aggregates age and the dynamics is non-ergodic. The implications for aging in RNA condensates and intrinsically disordered propteins are outlined.  Our study bridges the connection between the distinct morphologies in biomolecular aggregates and the expected range of complex aging dynamics. 

\section{Results}
\textbf{Word Complexity.} 
We consider sequences generated using three letters, A, B, and C. The letters combine to form words consisting of $n$ letters. We created the {\bf ABC} model that contains  letter words ($n$=3). The total number of such words is $3^3$ = 27. Examples of words are ABC, ACB, AAA, ABA, and so on. Following Herzel et. al.,~\cite{herzel1994entropies}, we define sequence complexity in terms of word entropy,
\begin{align} 
\begin{split}
\label{eq:H}
H_3=\sum_{i}^{}-p^{(n)}_i\log_2 p^{(n)}_i,
\end{split}  
\end{align}
where $p_i^{(n)}$, with $n$=3, is the probability of finding the combination of a word in a given sequence. A small (large) value of $H_3$ is a low (high) complexity sequence.  The minimum value of $H_3$, corresponding to the periodic arrangement of a word (ABCABC..), is $\log_2 3 \approx 1.58$ bits. The random arrangement of words yields the maximum value, $\log_2 3^3 \approx 4.75$ bits, corresponding to random sequences. The use of the word entropy to quantify sequence complexity  differs from the usage in the IDP literature. In the latter field, low complexity qualitatively refers to the frequency of occurrence of similar types of residues. For example, the N-terminal of the IDP, Fused In Sarcoma, is rich in QGSY (Glu-Gly-Ser-Tyr) repeats. The $H_3$ value for this sequence is 3.08 based on a specific amino acid classification scheme (see below).

\begin{figure*}[]
\centering
\includegraphics[width=0.8\textwidth]{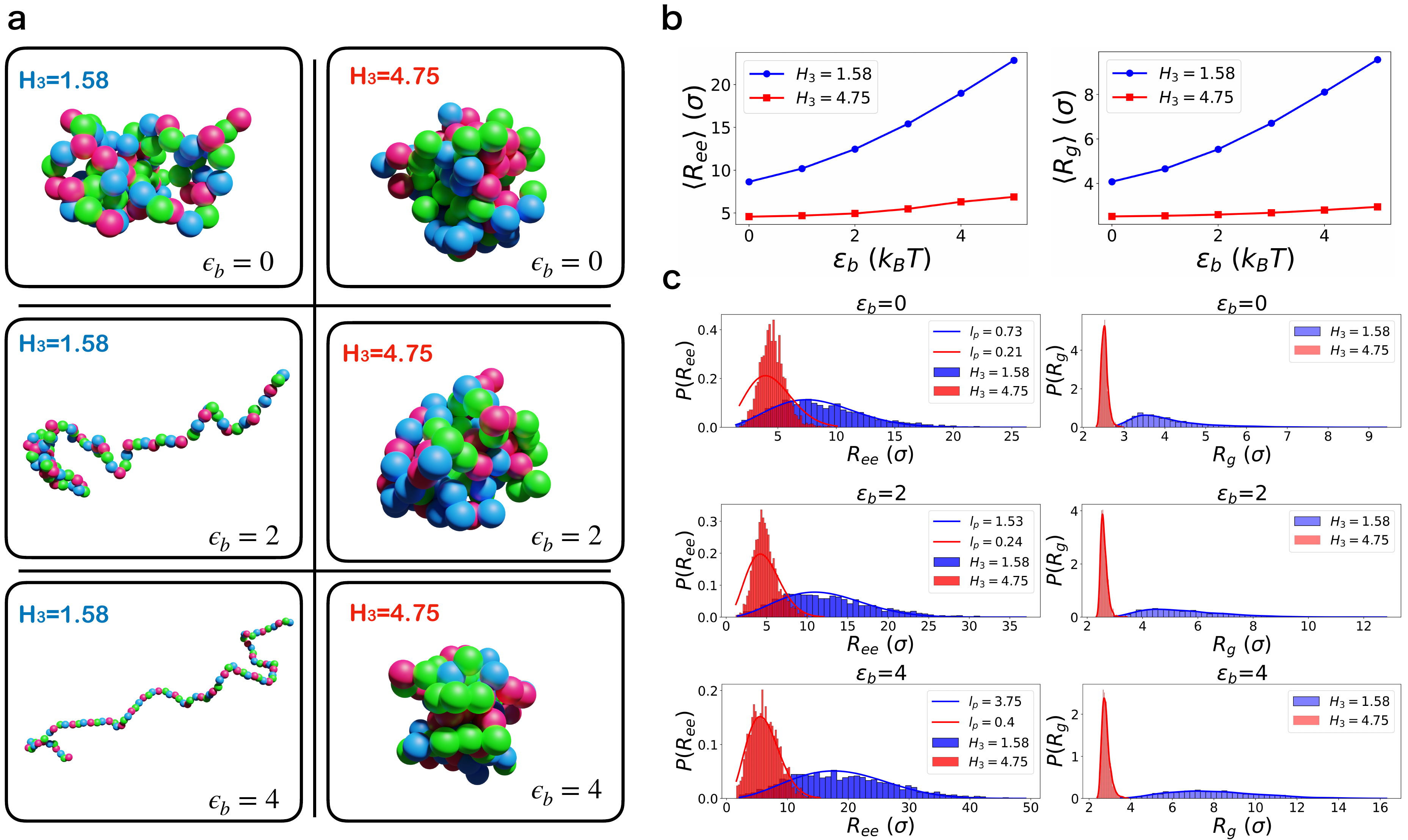}
\caption{\label{fig:Single_Morphology}{\bf Structural ensemble of protein monomers:} (a) Representative simulation snapshots. Bead colors denote the sequence letters, with red corresponding to A, green to B, and blue to C.
(b) Mean values of the end-to-end distance ($R_{ee}$) and radius of gyration ($R_g$) as a function of the bending rigidity, $\epsilon_b$. The blue and red curves represent $H_3=1.58$ and $H_3=4.75$, respectively. The unit length $\sigma$ is the bead diameter (Methods). 
(c) The distribution of the end-to-end distance (left) and radius of gyration (right) for $H_3=1.58$ (blue) and $H_3=4.75$ (red). The end-to-end distance distributions were fit using the analytical expression (Methods), thus allowing us to estimate the effective persistence length, $l_p\ (\sigma)$. The values of $l_p$ are $l_p=0.73$ and $l_p=0.21$ for $H_3=1.58$ and $H_3=4.75$, for $\epsilon_b=0$; $l_p=1.53$ and $l_p=0.24$ for $H_3=1.58$ and $H_3=4.75$, for $\epsilon_b=2$; and $l_p=3.75$ and $l_p=0.40$ for $H_3=1.58$ and $H_3=4.75$, for $\epsilon_b=4$. }
\end{figure*}

In our simulations, the letters are modeled as spherical beads, and a sequence refers to a particular arrangement of letters (see Fig.\ref{fig:Single_Morphology}a for an illustration). In the simplest realization of the model, identical letters have attractive interactions while interactions between non-identical letters are repulsive (see Methods for details).

\textbf{Monomer conformational ensembles depend on word entropy and rigidity.}
In addition to sequence complexity, the rigidity of the monomer not only controls the monomer structural ensemble but also plays a vital role in determining both the morphology and dynamics of the aggregates. 
In the {\bf ABC} model, the rigidity of the isolated monomer is determined by the angle bending potential (Methods) with strength $\epsilon_b$, which we measure in the unit of $k_BT$ where $k_B$ is the Boltzmann constant, and $T$ is the temperature. In the absence of the angle bending potential ($\epsilon_b$  = 0), the chain is flexible. For non-zero $\epsilon_b$, a rough estimate of the persistence length for homopolymer is $l_p \simeq l_b\epsilon_b/k_BT$ with $l_b$ being the bond length~\cite{midya2019phase}. 
Thus, in the {\bf ABC} model, the monomer is characterized by $H_3$ and $\epsilon_b$. 

Several striking features in Fig.~\ref{fig:Single_Morphology} are worth pointing out.  (i) Fig.~\ref{fig:Single_Morphology}a shows representative conformations for $H_3=1.58$ and $H_3=4.75$ for different bending rigidities. The conformations of the low and high-complexity sequences are markedly different. For a given $\epsilon_b$, the $H_3=1.58$ sequence, structures are considerably more expanded than the $H_3=4.75$ sequence.  
(ii) The mean values of the end-to-end distance ($R_{ee}$), as well as the radius of gyration, $R_g$ as a function of the bending rigidity, are plotted in Fig.~\ref{fig:Single_Morphology}b, and the corresponding distributions, $P(R_{ee})$ and $P(R_g)$, for $H_3=1.58$ and $H_3=4.75$ as a function of $\epsilon_b$ are shown in Fig.~\ref{fig:Single_Morphology}c.  The dependence of $R_{ee}$ and $R_g$ on $\epsilon_b$ of the two sequences is strikingly different.  Both $R_{ee}$ and $R_g$ increase significantly for the $H_3=1.58$ sequence but are relatively unchanged for the $H_3=4.75$ sequence. (iii) For a given $\epsilon_b$, both  $P(R_{ee})$ and $P(R_g)$ for the two sequences are also different.  The values of the peaks in the distributions are at much larger values of $R_{ee}$ or $R_g$ for the $H_3=1.58$ sequence than for the $H_3=4.75$  sequence. Furthermore, the widths of the distributions are much less for the high-complexity sequence relative to the low-complexity sequence. The distributions for the $H_3=1.58$ sequence sample a broad range of values. (iv) The distribution $P(R_{ee})$ for $H_3=1.58$ behaves like a polymer in a good solvent. We fit the $P(R_{ee})$ to the mean-field expression~\cite{hyeon2006kinetics} for semi-flexible chains, to calculate the effective persistence length $l_p$.  The fit for $H_3=1.58$ is excellent, while the $P(R_{ee})$ for $H_3=4.75$ deviates from the distribution for semi-flexible polymer, particularly at lower values of $\epsilon_b$. The effective persistence length obtained from the fit for $H_3=4.75$ is much smaller than for $H_3=1.58$. For instance, for $\epsilon_b=4$, $l_p=0.40(\sigma)$ and $l_p=3.75(\sigma)$ for $H_3=4.75$ and $H_3=1.58$, respectively. The small effective persistence length for the high complexity sequence brings together beads with favorable attractive interactions, leading to the compaction of the polymer.  On the other hand, the low complexity repeat sequence restricts the potential number of interactions in the monomer, leading to more extended conformations. We next investigated how these features in the monomer conformations affect the formation of aggregates. Because the monomer conformational ensemble of the two sequences is different, we expect that the dynamics and morphologies of the multi-chain systems have to vary dramatically. 

\subsection*{Interplay between complexity and rigidity on the morphologies of aggregates}
\begin{figure*}[]
\centering
\includegraphics[width=0.8\textwidth]{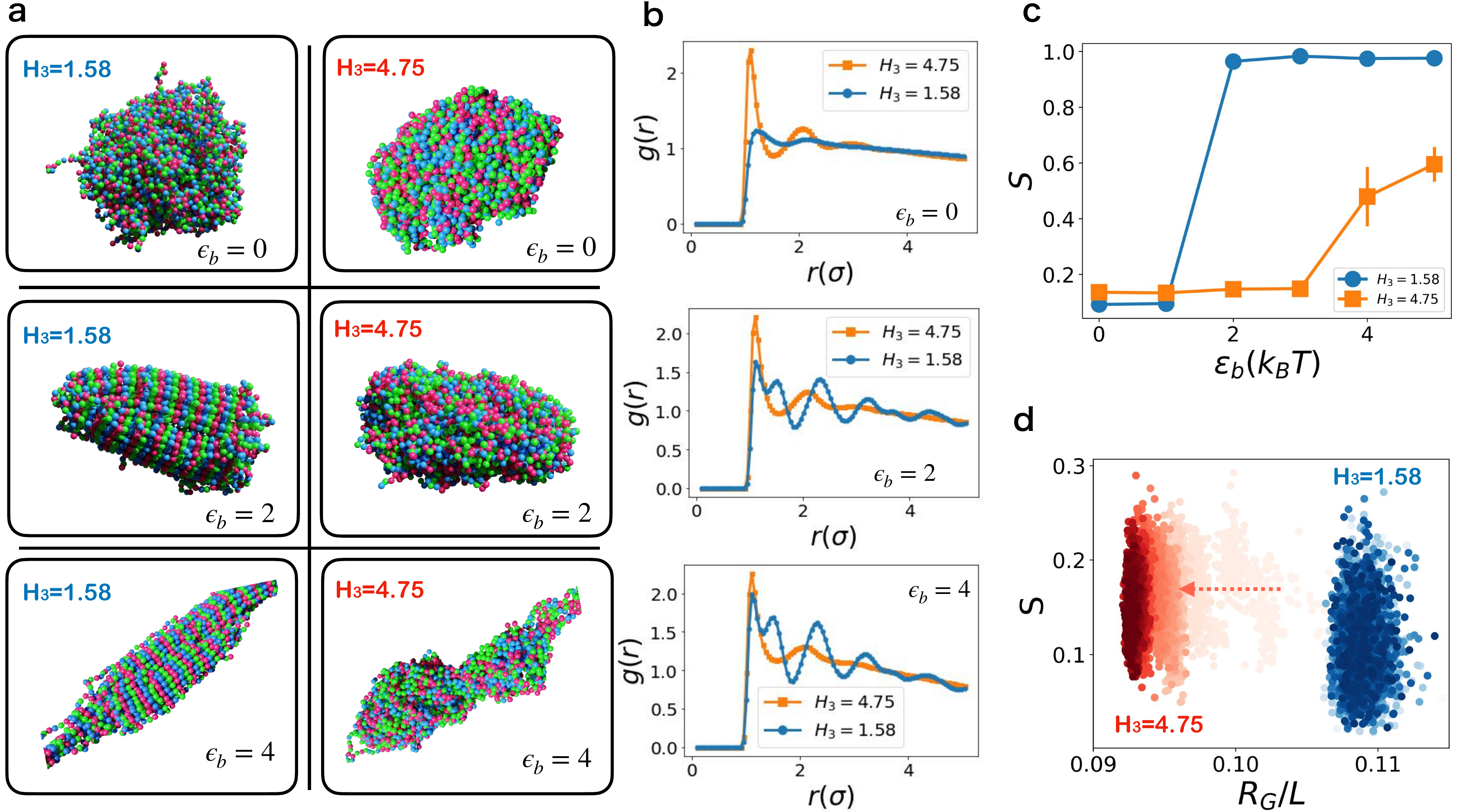}
\caption{\label{fig:Aggregates_Morphology}  {\bf Morphologies of protein aggregates:} (a) Representative simulation snapshots. Bead colors denote the sequence letters, with red corresponding to A, green to B, and blue to C.
(b) Radial distribution function, $g(r)$, with $r$ measured in bead diameter unit. Blue traces are associated with $H_3=1.58$, and red traces with $H_3=4.75$. The sequence from top to bottom corresponds to $\epsilon_b=0$, $\epsilon_b=2$, and $\epsilon_b=4$.
(c) Nematic order parameter, $S$, as a function of monomer rigidity, $\epsilon_b$. Blue (red) trace represents $H_3=1.58$ ($H_3=4.75$). Error bars reflect the standard error derived from six separate simulations (three unique sequences, each with identical $H_3$ and two ensembles or replicates per sequence).
(d) Trajectories illustrating the evolution of the morphology of aggregates ($\epsilon_b=0$). The $x$-axis represents the radius of gyration of aggregates, $R_G$, normalized by the length of the periodic box, $L=100\sigma$. The $y$-axis depicts the nematic order parameter, $S$. The opacity of the dots indicates progression in simulation time, with denser dots corresponding to more recent times. Blue and red dots signify $H_3=1.58$ and $H_3=4.75$, respectively. The white transparent dots correspond to the early times for $H_3=4.75$. The red arrow indicates the compaction of the aggregate over time for $H_3=4.75$.  The aggregate colored in blue undergoes small fluctuations whereas the high complexity aggregate in red visually conveys aging and shows the shrinkage, observed in the experiment ~\cite{jawerth2020protein}.  }
\end{figure*}

We first determined the structures in multi-chain simulations (see Methods for details of the model and simulations). Fig.\ref{fig:Aggregates_Morphology}a, shows representative snapshots for various parameter combinations: $H_3=1.58$, $H_3=4.75$, $\epsilon_b=0$, $\epsilon_b=2$, and $\epsilon_b=4$. For both high and low-complexity sequences, flexible monomer ensembles ($\epsilon_b=0$) lead to heterogeneous assembly.  In Fig.\ref{fig:Aggregates_Morphology}b, we display the radial distribution function, $g(r)$, as a function of the interparticle distance $r$, measured in the unit of the bead diameter, to analyze the extent of conformational order in the aggregates.  Although both sequences have liquid-like characteristics in the radial distribution, the $H_3=1.58$ sequence exhibits a longer range order at non-zero $\epsilon$. Interestingly,  at $\epsilon_b=0$ the low sequence complexity, $g(r)$ reveals no pronounced peaks, indicating the absence of long-range ordering within the aggregate. As $\epsilon_b$ increases, distinct peaks in $g(r)$ become evident, affirming the presence of structures with  short-range order.  For the $H_3=4.75$  sequence, $g(r)$ is essentially unchanged as $\epsilon_b$ varied. 
Although an increase in rigidity results in more extended aggregate shapes, it is not reflected in the $g(r)$. 
Intriguingly, an abrupt transition from disorder to order is evident as $\epsilon_b$ increases for the low complexity sequence. This is evident from the change in the nematic order parameter, $S$ on $\epsilon_b$ shown in Fig.\ref{fig:Aggregates_Morphology}c. There is an abrupt change in $S$ as $\epsilon_b=1$ increases from to $\epsilon_b=2$ in the $H_3=1.58$ sequence.  Conversely, the high complexity sequence displays resilience to ordering as rigidity increases. 

As a prelude to describing the dynamics in detail, we pictorially depict in Fig.\ref{fig:Aggregates_Morphology}d the evolution of a trajectory for an aggregate as a function of $S$ and and the radius of gyration (size) of the aggregate $R_G$ with $\epsilon_b=0$.  In the figure, lighter dots correspond to early times, while denser-colored dots represent late times. The trajectory for $H_3=4.75$ illustrates slow compaction of the aggregate over time (highlighted by the red arrow), which is strikingly different from the stationary fluctuations of the morphology observed for $H_3=1.58$. Similar figures for values of rigidity ($\epsilon_b$) are shown in the SI, section I.

\subsection*{Broken ergodicity and aging in aggregate dynamics}
It is difficult to distinguish between liquid and glass states based solely on the structure, as the conformations of both states are disordered ~\cite{RevModPhys.87.183,berthier2011theoretical}. A clear distinction emerges only upon investigating the dynamics, to which we shift our focus next.
To investigate the dynamics, we devised a measure to quantify the movements of the beads within the aggregates. To eliminate the overall translational and rotational diffusion within the aggregates, we define the distance between a pair of beads, as $R_{ij}(t)=|\bm{r}_{i}(t)-\bm{r}_j(t)|$, where $\bm{r}_{i}(t)$ ($\bm{r}_{j}(t)$) is the coordinate of bead $i$ ($j$) at time $t$. In terms of the displacement of the pairwise distance over the interval from $t_1$ to $t_2$,
\begin{equation} 
\label{eq:Rij}
\Delta R_{ij}(t_1,t_2)=R_{ij}(t_2)-R_{ij}(t_1),
\end{equation}
we define the mean squared pair displacement as, 
\begin{equation} 
\label{eq:MSD}
\langle \Delta R^2(t_1,t_2) \rangle \equiv  \Big\langle \frac{1}{N_{p}}  \sum_{(ij)}\big(\Delta R_{ij}(t_1,t_2)\big )^2 \Big\rangle. 
\end{equation}
The above equation measures the average displacement of two beads in the time interval $(t_2 - t_1)$. In Eq. \ref{eq:MSD}, $N_p$ is the number of pairs, $(ij)$ designates a pair of beads within the aggregate, and $\langle \cdots \rangle$ is the ensemble average. We exclude adjacent bead pairs in the monomers from the pair count but include both pairs that are within and between polymer chains.  It is important to note that the 2-point function in Eq.\ref{eq:MSD}  is defined for bead pairs within the aggregate, thus differing from the conventional mean squared displacement, which is defined for the displacement of individual beads. We refer to Eq. \ref{eq:MSD} as pMSD (pair Mean Squared Displacement) to highlight the difference.

By tuning $H_3$ and $\epsilon_b$ in the {\bf ABC}  model, one can generate various phases (Fig. \ref{fig:Aggregates_Morphology}). The nematic order parameter and the radial distribution functions allow us to distinguish between the morphologies adopted by the aggregates. Ordered states of the low complexity sequence at high $\epsilon_b$ (bottom panels in Fig. \ref{fig:Aggregates_Morphology}a) could be characterized by nematic order parameter (Fig. \ref{fig:Aggregates_Morphology}c). However, disordered liquid and glass-like states usually have similar structures. But in contrast to liquids, glassy materials often exhibit age-dependent dynamics, in which the material properties vary based on the measurement time.   The absence of aging would imply liquid-like ergodic behavior. Thus, measures of aging could be used to discern the differences between glass and liquid. 

\begin{figure}[]
\centering
\includegraphics[width=0.5\textwidth]{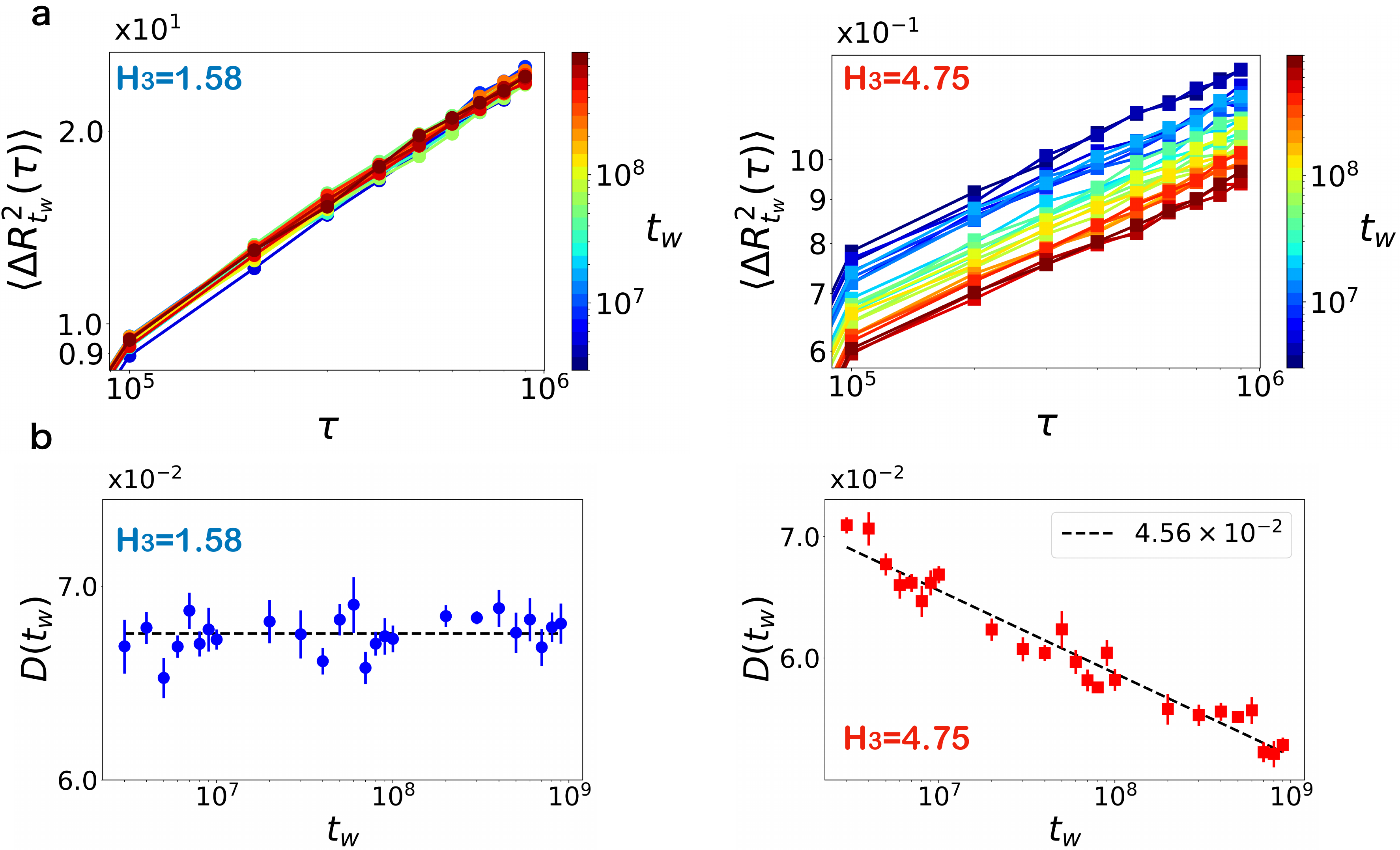}
\caption{\label{fig:MSD} {\bf Age-dependent pair mean squared displacement for $\bm{\epsilon_b=0}$:} (a) pMSD  for low complexity sequence ($H_3=1.58$, left) and high complexity sequence ($H_3=4.75$, right). The color gradient shows the age ($t_w$) of the system. In contrast to the low-complexity sequence, the pMSD for the high-complexity sequence depends on the waiting time, $t_w$.
(b) The generalized diffusion coefficient $D(t_w)$ as a function of the age $t_w$ obtained using Eq.(\ref{eq:GDC}). The error bars are the root mean squared error of the fit for $D(t_w)$. The dashed line is a power-law fit with exponent $\alpha=(4.56 \pm 0.46) \times 10^{-2}$ for $H_3=4.75$. For $H_3=1.58$, the fit is independent of $t_w$. The error of the exponent is $95 \%$ confidence interval of the parameter estimate. The figures on the left and right correspond to $H_3=1.58$ and $H_3=4.75$, respectively.  
The ensemble average is performed over $10$ shifted time windows between the successive $t_w$s and $2$ different trajectories, generating $20$ realizations. 
}
\end{figure}

If aging is relevant then the material properties would depend on both the observation time, $\tau$, and the waiting time (age), $t_w$, the time at which measurement begins. The age of the aggregate and observation time are related by, $t = t_w + \tau$. 
By choosing $t_2=t_w+\tau$ and $t_1=t_w $ in Eq.(\ref{eq:MSD}), we define the age-dependent pMSD, 
\begin{equation} 
\label{eq:MSD_tw}
\langle \Delta R_{t_w}^2(\tau) \rangle \equiv  \langle \Delta R^2(t_w+\tau,t_w) \rangle. 
\end{equation}
If $t_w$ and $\tau$ are well separated, we expect that pMSD should factorize as,  
\begin{equation} 
\label{eq:GDC}
\langle \Delta R_{t_w}^2(\tau) \rangle= D(t_w)\tau ^ \kappa \sim t_w^{-\alpha} \tau ^ \kappa,
\end{equation}
where \(D(t_w)\) is the age-dependent generalized diffusion coefficient, and \(\kappa\), which is expected to differ from unity, is an anamolous exponent. When the aggregate ages, we expect the generalized diffusion coefficient should decrease as a power-law, $D(t_w) \sim (t_w)^{-\alpha}$ with an exponent $\alpha$.
Fig.\ref{fig:MSD}a, showing $\langle \Delta R_{t_w}^2(\tau) \rangle$ as a function of $\tau$ for $\epsilon_b=0$, confirms that this is indeed the case. From the power-law relation between $\langle \Delta R_{t_w}^2(\tau) \rangle$ and $\tau$ in Fig.\ref{fig:MSD}a, we find \(\kappa\) values of \(0.43\) and \(0.21\), for $H_3=1.58$ and $H_3=4.75$, respectively. The fractional values of $\kappa$ for both the sequences imply that the pMSD dynamics is subdiffusive.

The generalized diffusion coefficient \(D(t_w)\) are different for low and high-complexity sequences. For $H_3=1.58$, \(D(t_w)\) is age-independent, as evidenced in Fig.\ref{fig:MSD}b, left panel. This implies the aggregate under this condition is ergodic. In contrast, \(D(t_w)\) for $H_3=4.75$ is clearly age-dependent, with a power-law fit in \(t_w\) yielding an exponent \(\alpha=(4.56 \pm 0.46)\times 10^{-2}\). Consequently, the diffusivity of beads within the aggregates decreases and is accompanied by aging in high-complexity sequences. These contrasting behaviors of \(D(t_w)\) for $\epsilon_b = 0$ between low and high complexity sequences illustrate the fluid-like and glassy nature of the aggregates, respectively. We show the pMSD for non-zero $\epsilon_b$ in section II of the SI. 

\subsection*{Dynamics of the structural overlap}
\begin{figure}[]
\centering
\includegraphics[width=0.45\textwidth]{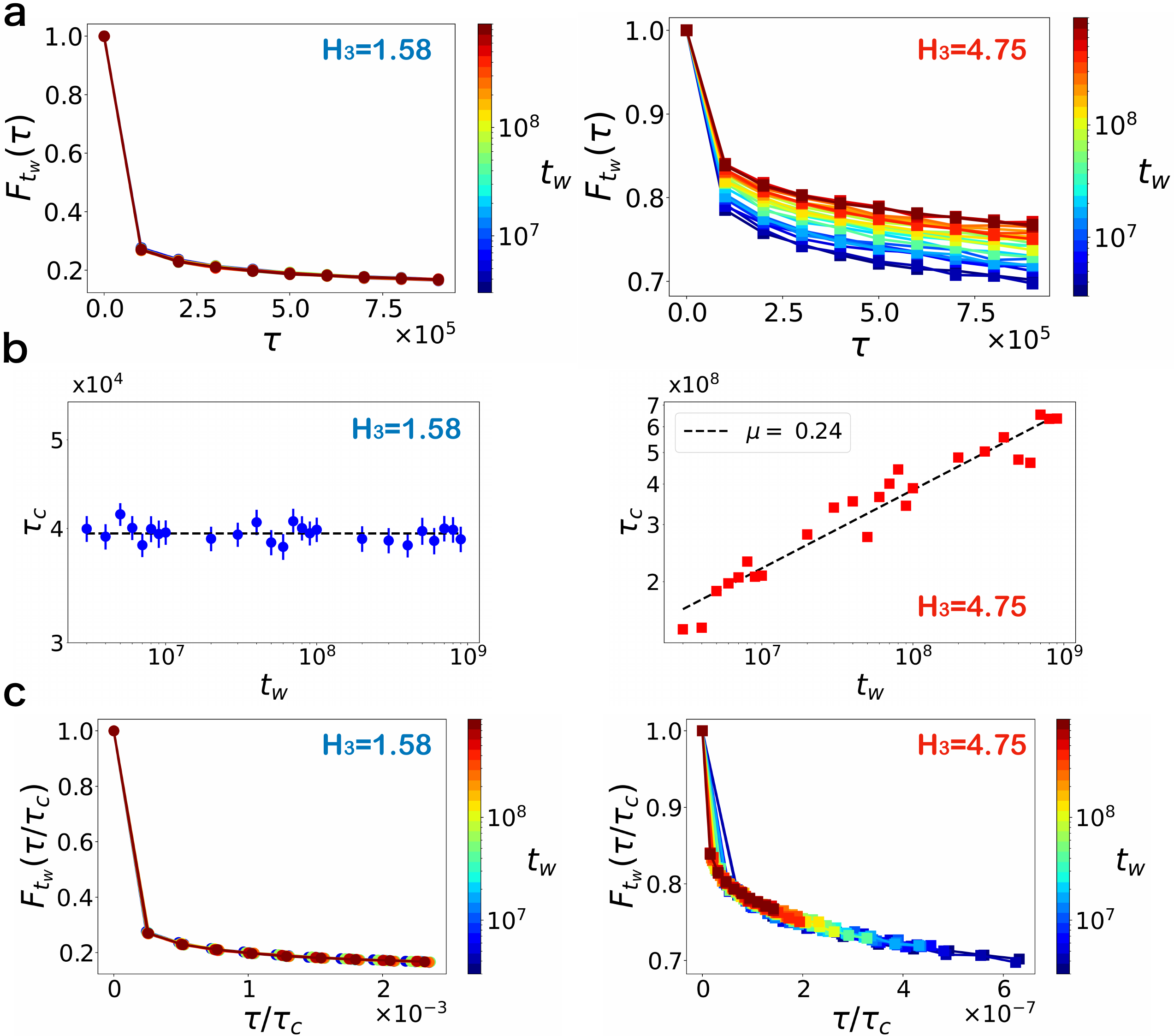}
\caption{\label{fig:Dynamics_epb=0} {\bf Structural relaxation in aggregates for $\bm{\epsilon_b=0}$:} (a) Comparison of the relaxation of $F_{t_w}(\tau)$ for the low complexity sequence ($H_3=1.58$, left) with the high complexity sequence ($H_3=4.75$, right). The color gradients, with the scales on the right of each figure, show the age of the system ($t_w$).
(b) Characteristic relaxation time $\tau_c$ as a function of the ageing time, $t_w$. The time constant, $\tau_c$, is obtained by fitting Eq. (\ref{stretch}) to $F_{t_w}(\tau)$ with $\beta=0.20$. The error bars of the fit (root mean squared error) for $H_3=4.75$ are smaller than the markers.  The dashed line is a power-law fit, $\tau_c \sim (t_w)^{\mu}$, with an exponent $\mu=0.24\pm 0.03$ for $H_3=4.75$. The error of the exponent is the $95 \%$ confidence interval of the estimate. For $H_3=1.58$, $\tau_c$ is independent of $t_w$. The figures on the left and right correspond to $H_3=1.58$ and $H_3=4.75$, respectively. 
(c) Collapse of $F_{t_w}(\tau)$ by rescaling $\tau$ by $\tau_c$. The left and right figures represent $H_3=1.58$ and $H_3=4.75$, respectively.  
}
\end{figure}

The relaxation behavior of the aggregates may be assessed by examining the decay of the structural correlation function, which is the analog of the overlap introduced in the context of spin glasses and protein folding~\cite{guo1995kinetics,RevModPhys.87.183}. In terms of $\Delta R_{ij}(t_1,t_2)=R_{ij}(t_2)-R_{ij}(t_1)$, the overlap function is, 
\begin{equation} 
\label{eq:F}
F(t_1,t_2) \equiv \Big \langle \frac{1}{N_p}\sum_{(ij)}\Theta(r_c-\Delta R_{ij}(t_1,t_2)) \Big \rangle,
\end{equation}
where $\Theta(x)$ is the Heaviside step function: $1$ for $x\geq 0$ and $0$ for $x<0$, and $r_c$ is a cut off distance that is set to $\sigma$. Similar to the pMSD, we are interested in the age-dependent relaxation of the structural correlation, which is given by,
\begin{equation} 
\label{eq:}
F_{t_w}(\tau) \equiv  F(t_w+\tau,t_w). 
\end{equation}
In Fig.\ref{fig:Dynamics_epb=0}a, we plot $F_{t_w}(\tau)$ for aggregates composed of flexible monomers ($\epsilon_b=0$). The variation in the decay over $t_w$ is more pronounced for $H_3=4.75$ sequence than for $H_3=1.58$ sequence. The relaxation curves for both sequences are well fit by the stretched exponential function,
\begin{equation}
 F_{t_w}(\tau) = e^{-(\tau/\tau_c)^{\beta}},
 \label{stretch}
 \end{equation}
with small stretching exponent $\beta=0.20$.  

The results in Fig.\ref{fig:Dynamics_epb=0} lead to the following conclusions. (1) In contrast to the $H_3=4.75$ sequence, we find that $F_{t_w}(\tau)$ for the $H_3=1.58$ sequence decays rapidly. On the time scale $\tau \approx 8 \times 10^5$  the structural overlap function decreases to $F_{t_w}(\tau) \approx 0.2$  for $H_3=1.58$ whereas for the $H_3=4.75$ sequence the decay is much less ($F_{t_w}(\tau) \approx$ 0.7 - compare the two panels in Fig.\ref{fig:Dynamics_epb=0}a). The decay of $F_{t_w}(\tau)$ depends on the aging time for $H_3=4.75$ whereas it is independent of $t_w$ for $H_3=1.58$. (2) More importantly, the relaxation time, $\tau_c$, does not depend on the waiting time ($t_w$) for the low complexity sequence, which implies that aggregate is ergodic. In sharp contrast, $\tau_c$ increases as a power law,   $\tau_c \sim (t_w)^{\mu}$ ($\mu \approx$ 0.24) for $H_3=4.75$ sequence. This implies that the aggregates of the high-complexity sequence ages, giving rise to non-ergodic behavior. The power-law increase in $\tau_c$ as the system $t_w$ increases is a hallmark of aging in glassy systems~\cite{struik1977physical,PRXLife.1.013006,berthier2011theoretical}. (3) Strikingly,  Fig.\ref{fig:Dynamics_epb=0}c shows the relaxation curves for both high and low complexity sequences collapse onto a single master curve by rescaling $\tau$ by $\tau_c$, indicating self-similarity in the structural relaxation mechanism across different aggregate ages. Analysis of the dynamics within aggregates emphasizes the differences between protein aggregates resembling liquid-like and glassy states. Such differences are challenging to discern from morphology alone, suggesting that aging measurements are needed to understand dynamics in condensates.


\subsection*{Bending rigidity has a dramatic effect on the aging dynamics}
\begin{figure}[]
\centering
\includegraphics[width=0.5\textwidth]{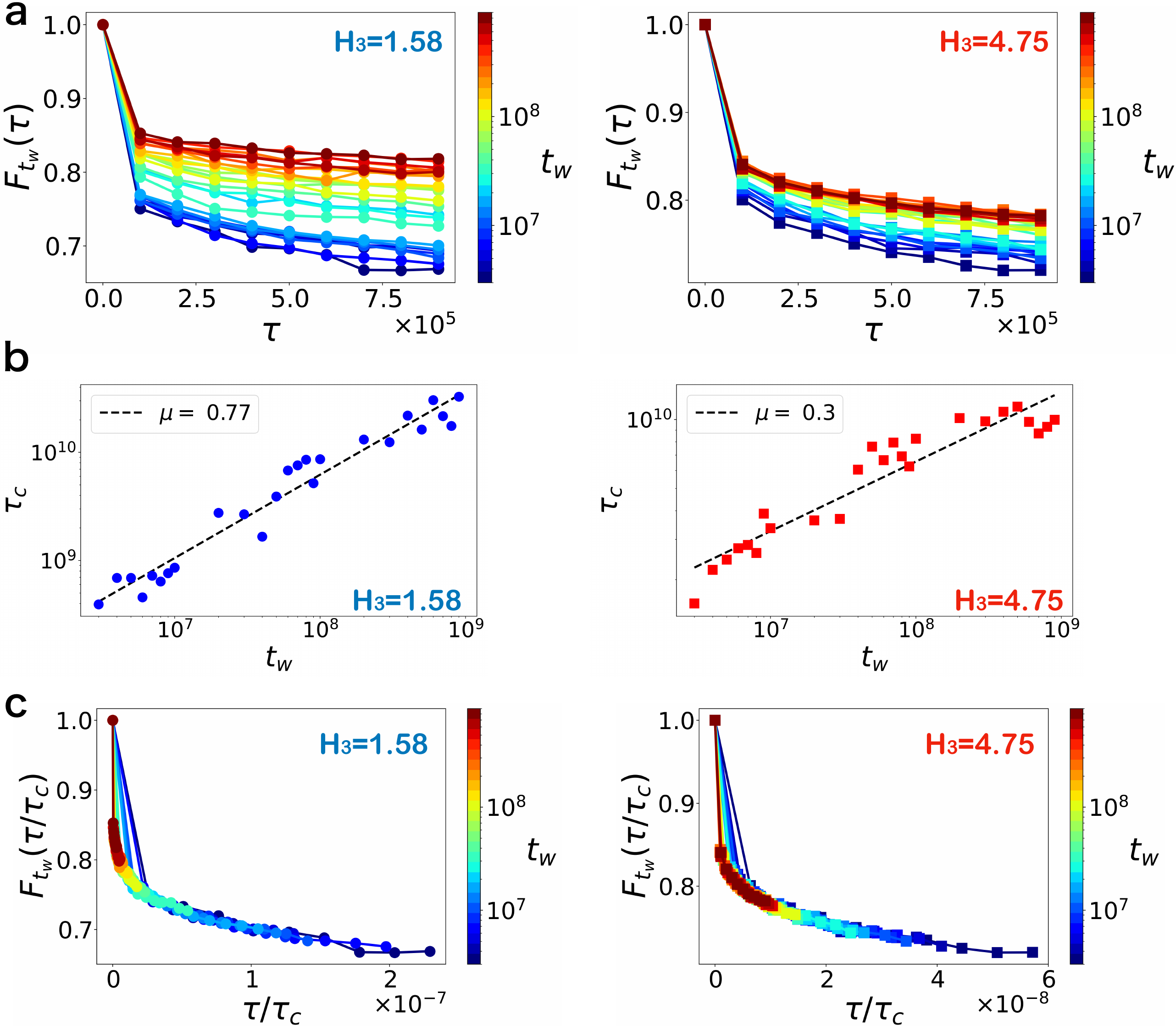}
\caption{\label{fig:Dynamics_epb=2} {\bf Structural relaxation of protein aggregates for $\bm{\epsilon_b=2}$:} (a) Structural relaxation $F_{t_w}(\tau)$ for low complexity sequence ($H_3=1.58$, left) compared to high complexity sequence ($H_3=4.75$, right). The color gradient indicates the system's age ($t_w$).
(b) Characteristic relaxation time $\tau_c$ as a function of the age $t_w$. The characteristic relaxation time $\tau_c$ is derived from fitting Eq. (\ref{stretch}) to $F_{t_w}(\tau)$ with $\beta=0.15$. The error bars (root mean squared error) are smaller than the symbols. The dashed line is a power-law fit, $\tau_c \sim (t_w)^\mu$, with exponent $\mu=0.77\pm 0.08$ for $H_3=1.58$ and $\mu=0.30 \pm 0.05$ for $H_3=4.75$. The errors of the exponents are the $95 \%$ confidence interval of the estimates.
(c) Collapse of $F_{t_w}(\tau)$ by rescaling $\tau$ with $\tau_c$. The left and right figures represent $H_3=1.58$ and $H_3=4.75$, respectively.  }
\end{figure}

\begin{figure}[]
\centering
\includegraphics[width=0.5\textwidth]{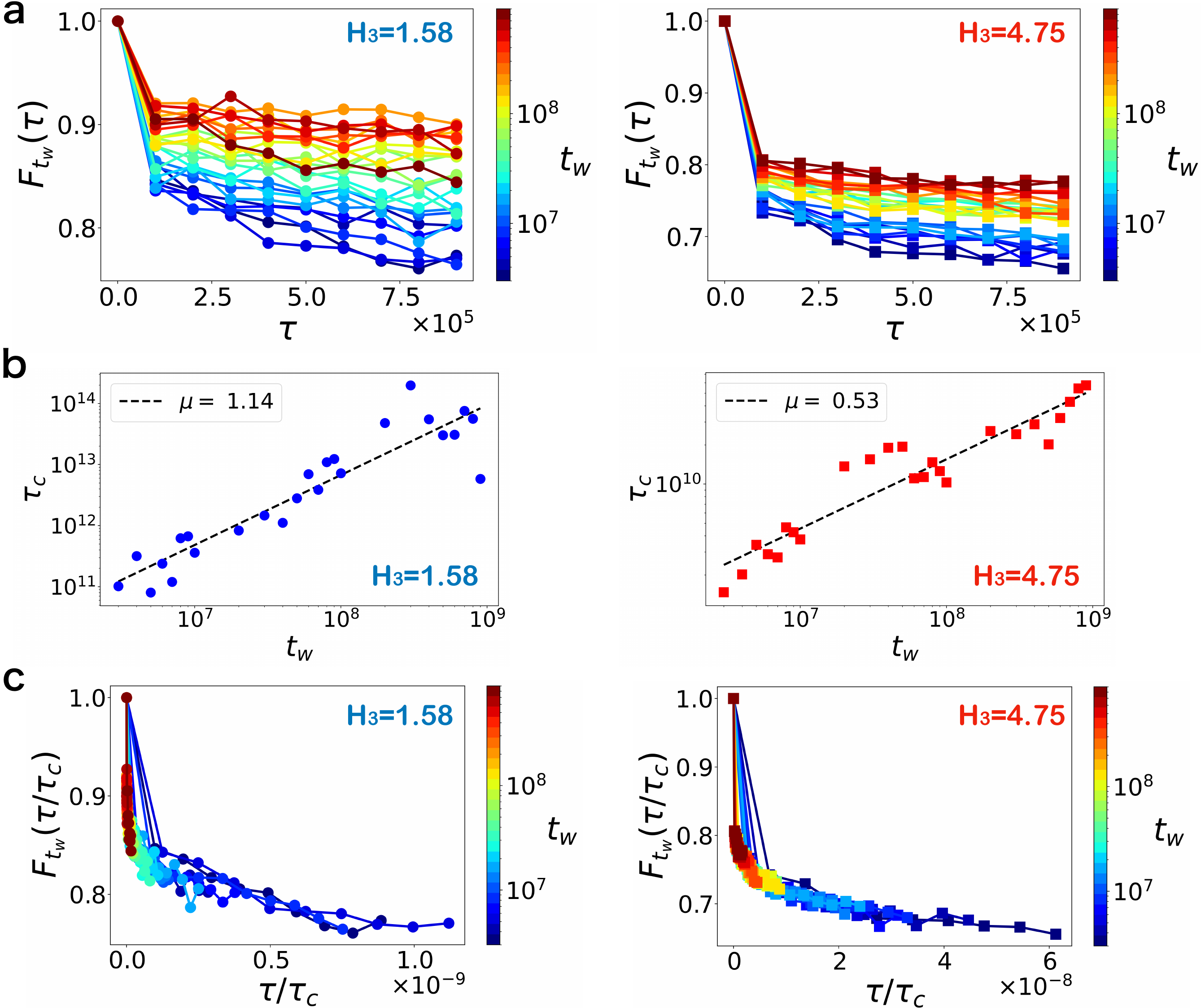}
\caption{\label{fig:Dynamics_epb=4} {\bf Structural relaxation of protein aggregates for $\bm{\epsilon_b=4}$:} (a) Structural relaxation $F_{t_w}(\tau)$ for low complexity sequence ($H_3=1.58$, left) compared to high complexity sequence ($H_3=4.75$, right). The color gradient indicates the system's age ($t_w$).
(b) Characteristic relaxation time $\tau_c$ as a function of the age $t_w$. The characteristic relaxation time $\tau_c$ is derived from fitting Eq. (\ref{stretch}) to $F_{t_w}(\tau)$ with $\beta=0.12$. The error bars (root mean squared error), obtained from the fits, are smaller than the symbols. The dashed line is a power-law fit, $\tau_c \sim (t_w)^\mu$, with exponent $\mu=1.14 \pm 0.21$ for $H_3=1.58$ and $\mu=0.53 \pm 0.08$ for $H_3=4.75$. The errors of the exponents are the $95\%$ confidence interval of the estimates.
(c) Collapse of $F_{t_w}(\tau)$ by rescaling $\tau$ with $\tau_c$. The left and right figures represent $H_3=1.58$ and $H_3=4.75$, respectively.}
\end{figure}

In Fig.\ref{fig:Dynamics_epb=2} and Fig.\ref{fig:Dynamics_epb=4}, we present the analysis of $F_{t_w}(\tau)$ at $\epsilon_b=2$ and $\epsilon_b=4$, respectively.  There are several important differences between the dynamics of the flexible and stiff chains.  First, $F_{t_w}(\tau)$ decays slowly, exhibiting extreme glassy behavior. This is reflected in the change in the stretching exponent $\beta$, which decreases as the rigidity increases, going from $\beta=0.20$ ($\epsilon_b=0$) to $\beta=0.12$  ($\epsilon_b=4$).  The decrease of $\beta$ shows that the aggregates become more solid-like as the bending rigidity is increased. Second, the aggregates of the low-complexity sequence at $\epsilon_b \ne 0$ have appreciable nematic order, in contrast to the equilibrium liquid-like behavior for $\epsilon_b=0$ (compare Fig. \ref{fig:Dynamics_epb=0}a and Fig. \ref{fig:Dynamics_epb=2}-\ref{fig:Dynamics_epb=4}a). The exponent characterizing the power law relation, $\tau_c \sim (t_w)^{\mu}$, increases as $\epsilon_b$ increases. The $\mu$ value is larger for the $H_3=1.58$ sequence than the $H_3=4.75$ sequence. Intriguingly, the relaxation time for $\epsilon_b=4$ and $H_3=1.58$ sequence (Fig.\ref{fig:Dynamics_epb=4}b) is large and exceeds $\sim 10^{13}$, which is in accord with the slow dynamics in the formation of fibril structure (Fig. \ref{fig:Aggregates_Morphology}).  Third, although aging dynamics in both the low and high complexity sequences are qualitatively similar (Fig.\ref{fig:Dynamics_epb=2}-\ref{fig:Dynamics_epb=4}), there are major quantitative differences. The relaxation time, $\tau_c$ for $H_3=1.58$ is over three orders of magnitude greater than for $H_3=4.75$ (Fig.\ref{fig:Dynamics_epb=4}b), which is substantial given the finite number of polymer chains simulated here.  

The origin of the large difference may be understood in terms of the morphologies of the aggregates of the sequences (Fig. \ref{fig:Aggregates_Morphology}). The ground state of the $H_3=1.58$, at $\epsilon_b \ne 0$, is an ordered low entropy fibril with high nematic order (Fig. \ref{fig:Aggregates_Morphology}a and b). Therefore, the ordered state is reached, fluctuations away from the state are unlikely.  In contrast, the $H_3=4.75$ sequence is disordered and exhibits glass-like dynamics. Fourth, like in Fig. \ref{fig:Dynamics_epb=0}, $F_{t_w}(\tau)$ curves superimpose upon scaling by the global relaxation time, $\tau_c$, which suggests that the aging mechanism in protein aggregates is universal, which reflects the underlying glassy dynamics. Slow glass-like dynamics arise because the aggregates are trapped in one of several distinct metastable states. The transition from one long-lived metastable state to another would occur by activated processes, which is the physical basis of the theory of glass transition~\cite{kirkpatrick1989random}.  The trap model~\cite{monthus1996models,bouchaud1992weak}, used to explain aging in protein condensates~\cite{takaki2023theory}, also assumes that, with increasing time, the system can explore deeper and deeper minima by overcoming barriers in the rugged energy landscape. 

\subsection*{Protein aggregates are dynamically heterogeneous}
Glassy materials are spatially heterogeneous and characterized by persistent variations in the dynamics between different regions in an otherwise homogeneous system, a phenomenon commonly known as dynamical heterogeneity in the study of structural glasses~\cite{RevModPhys.87.183,berthier2011theoretical}. Dynamic heterogeneity may be quantitatively analyzed using the variance of the overlap function. This measure was introduced to distinguish between paramagnetic and spin glass phases in spin glasses without inversion symmetry (see Eq. 2.10 in~\cite{kirkpatrick1988comparison}), and has been more thoroughly explored in the study of structural glasses \cite{Toninelli05PRE,Donati02JNon-CrystSolids}. The fourth-order susceptibility is defined as, 
\begin{equation} 
\label{eq:chi4}
\chi_4(\tau) \equiv N_p \big \langle \big(F_{}(\tau) - \langle F_{}(\tau) \rangle \big )^2 \big \rangle. 
\end{equation}
It should be noted that unlike in simulations of glass-forming systems in which single particle MSD is used for the overlap function, here we use pMSD (Eq.\ref{eq:MSD}), the analog of the order parameter  in protein folding~\cite{guo1995kinetics}. 

\begin{figure}[]
\centering
\includegraphics[width=0.4\textwidth]{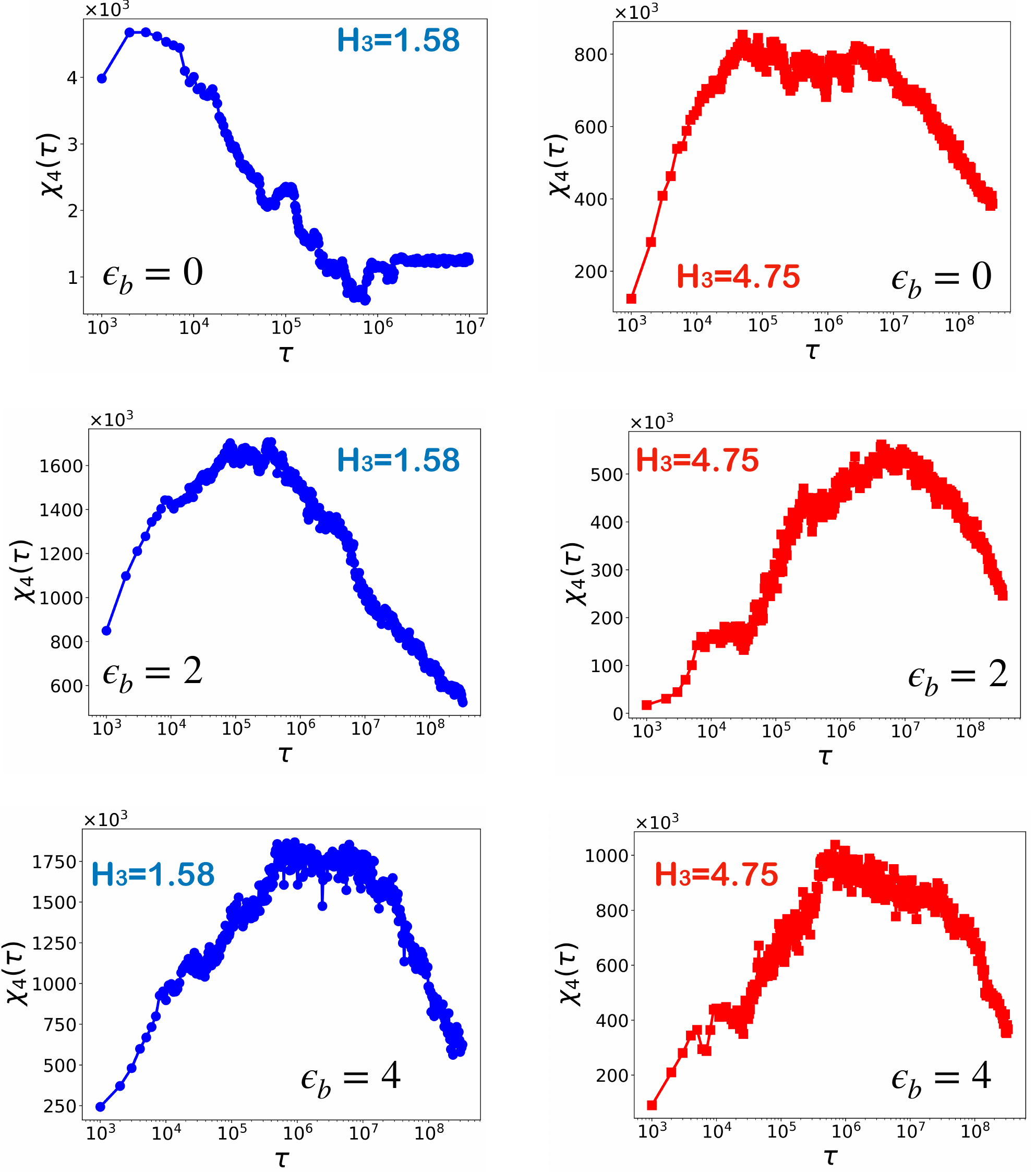}
\caption{\label{fig:chai4} {\bf Dynamical heterogeneity in protein aggregates:}
The left panels, in blue, depict $\chi_4(\tau)$ for $H_3=1.58$, while the right panels correspond to $H_3=4.75$. Arranged from top to bottom, the figures correspond to $\epsilon_b=0$, $\epsilon_b=2$, and $\epsilon_b=4$, respectively. For $H_3=1.58$ and $\epsilon_b=0$, the trajectory was divided into 100 blocks to calculate the ensemble needed for computing the variance [Eq.(\ref{eq:chi4})], ensuring that the length of $\tau$ spans the peak position of $\chi_4(\tau)$. For the other parameter sets, the trajectory was segmented into 3 blocks. Additionally, 2 independent trajectories were used, resulting in 200 ensemble sets for $H_3=1.58$ and $\epsilon_b=0$, and 6 ensemble sets for the other parameters. }
\end{figure}

Because dynamical heterogeneity in glassy materials manifests over a long timescale relative to $\tau$, 
we extended the observation time $\tau$ to include the peaks in $\chi_4(\tau)$. The peak positions and magnitudes in $\chi_4(\tau)$ reveal the timing and the extent of correlated motion of the beads within the aggregates. Fig.~\ref{fig:chai4} shows $\chi_4(t)$ for various $\epsilon_b$ and $H_3$ values. The peak position in $\chi_4(\tau)$ occurs at short times and the associated maximum is not pronounced for $\epsilon_b=0$ and $H_3=1.58$, which is consistent with the absence of aging for this system. The profiles for other parameters exhibit the characteristic peaks in $\chi_4(\tau)$ - a hallmark of glassy systems exhibiting dynamical heterogeneity.  As $\epsilon_b$ increases with $H_3=1.58$, both the peak positions and the magnitudes of $\chi(\tau)$ shift to larger values, indicating an expansion in the time and region of correlated motions. In contrast, the peak positions and magnitudes of $\chi_4(\tau)$ for $H_3=4.75$ are relatively constant across different $\epsilon_b$ values, signifying that high complexity sequences are less affected by increases in rigidity.  Interestingly, except for the $\epsilon_b=0$ and $H_3=1.58$, all other aggregates have glass-like or solid-like characteristics, consistent with Fig.\ref{fig:Dynamics_epb=2}-\ref{fig:Dynamics_epb=4}. 

Dynamical heterogeneity, characterized by the coexistence of mobile (fluid-like) and immobile (solid-like) regions within aggregates, has indeed been recently observed experimentally in FUS protein condensates~\cite{shen2023liquid}.  Our study shows that dynamical heterogeneity might be a defining characteristic of solidification and aging in biological condensates.

\section{Discussion}
\subsection*{Aging protein condensates and Maxwell glass}
Recent experiments suggest that the solidification of protein condensates and vitrification of materials may share common mechanisms~\cite{winter2017solidification,jawerth2020protein}.
Passive rheology experiments~\cite{jawerth2020protein} on a few proteins showed that the diffusion coefficients of the tracer beads within the protein condensates varied with the age of the material. On the experimental time scale $\tau/t_w \ll 1$, the dynamics are predominantly diffusive, with the mean squared displacement being proportional to the observation time, $\tau$.  A recent theoretical study~\cite{takaki2023theory} related the standard single-particle mean squared displacement and the aging,
\begin{equation} 
\label{eq:}
\frac{d}{dt} \langle \Delta x^2 \rangle (t) \sim t^{\mu - 1},
\end{equation}
where $\langle \Delta  x^2 \rangle (t)$ is the mean squared displacement of the tracer, and the exponent $\mu$ ranges from zero to unity. For  $\tau \ll t_w$, the age-dependent diffusion coefficient is, 
\begin{equation} 
\label{eq:}
D(t_w)   \sim t_w^{\mu - 1} \sim t_w^{-\alpha}.
\end{equation}
This relation predicts that the diffusion coefficient becomes decreases as the condensates age. 
In the current study, we computed the pMSD [Eq.(\ref{eq:MSD})] to quantify the dynamics of beads within the aggregates. We found that the generalized diffusion coefficient [Eq.(\ref{eq:GDC})] for high complexity sequences diminishes as $t_w$ increases,
which accords with the experiments studies of aging protein condensates~\cite{jawerth2020protein}. The pMSD exhibits scalies as $\tau^{\kappa}$ where $\kappa < 1$ (specifically, $\kappa=0.21$ for $H_3=4.75$), in contrast to the $\kappa=1$ that is expected for diffusive motion. The sub-diffusive behavior in pMSD may arise from the polymeric nature of the chains in the aggregates, where the constituent pair of beads are correlated, in contrast to the tracer particles in the experiments.
For further discussion about the connection between the trap model and the simulation, see SI section III. 

The aging exponent $\mu=1.14$ for $H_3=1.58$ and $\epsilon_b=4$ exceeds unity (see Fig.\ref{fig:Dynamics_epb=4}). This combination of the parameters ($H_3$ and  $\epsilon_b$) falls in the fibril-forming regime, (see Fig.\ref{fig:Aggregates_Morphology}a). Theoretical models of condensate aging~\cite{bouchaud1992weak,monthus1996models} that do not incorporate polymer rigidity predict an aging exponent of less than or equal to $1$~\cite{takaki2023theory,lin2022modeling}. However, experimentally, aging exponents exceeding unity have been observed~\cite{jawerth2020protein}. Our simulations suggest that the rigidity and structure formation in aggregates could be the neason for these larger exponents.  Such an interpretation is consistent experiments~\cite{jawerth2020protein} which have reported  fibril formation.


\subsection*{Connections to Intrinsically Disordered Proteins (IDPs)}
\begin{figure*}[]
\centering
\includegraphics[width=\textwidth]{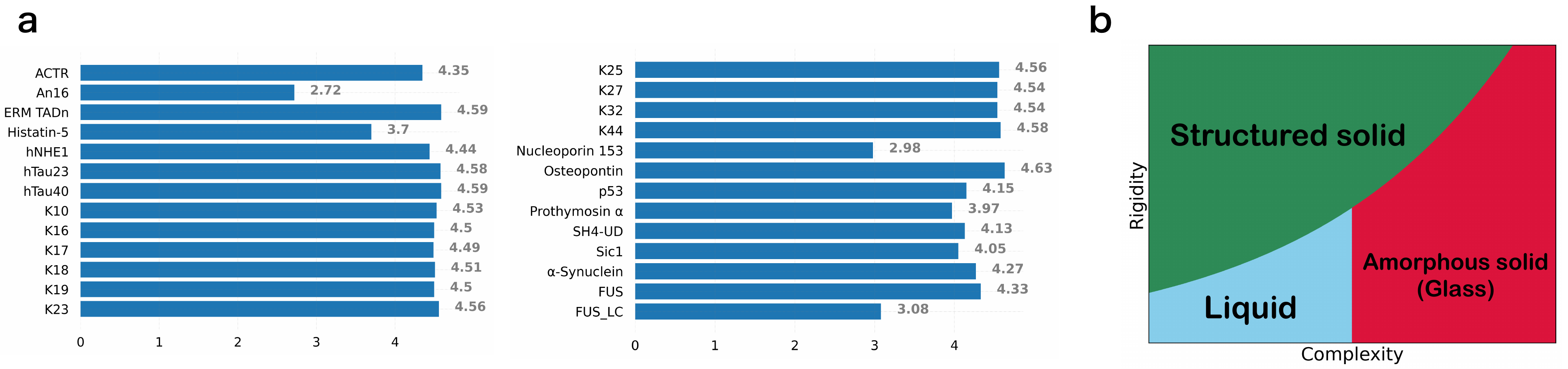}
\caption{\label{fig:State_Diagram}(a) Complexity of sequence as defined in Eq.(\ref{eq:H}), calculated for a range of intrinsically disordered proteins. The displayed numerical values are the $H_3$ values for the proteins. Protein sequences, adapted from Ref.~\cite{baul2019sequence}, are converted into three distinct types of amino acids as detailed in the main text.
(b) Conceptual state diagram of protein aggregates derived from our study. The x-axis is complexity and y-axis is the rigidity of the constitutive monomers in the aggregates.}
\end{figure*}

We focused on the extremes of sequence complexity ($H_3=1.58$ and $H_3=4.75$) by considering a three-word {\bf ABC} model. Real proteins exhibit a spectrum of sequence complexities, which should display a broad range of behavior. First, we calculated $H_3$ for IDPs by classifying amino acids into three groups, which are hydrophobic (A,I,L,F,V,P,G), polar (Q,N,H,S,T,C,W,Y,M), and charged (R,K,D,E). One-letter code for amino acids used. Based on this classification, we calculated $H_3$ for 26 IDPs, which are likely to form protein condensates~\cite{uversky2017intrinsically}.  The word entropy for these sequences is in the range $2.7 \le H_3 \le 4.6$ (Fig.\ref{fig:State_Diagram}a).  Previously we showed~\cite{baul2019sequence} that almost all of the IDPs, listed in Fig.\ref{fig:State_Diagram}a, behave as polymers in good solvents under conditions used in Small Angle X-ray scattering experiments. It follows that the IDPs listed in Fig.\ref{fig:State_Diagram}a are flexible.     In light of the findings in Fig.\ref{fig:Dynamics_epb=0}-Fig.\ref{fig:Dynamics_epb=4}, we predict that they are likely to adopt highly heterogeneous morphology.  Notably, the $H_3$ value of the low complexity domain of FUS, denoted as FUS\_LC, is sufficiently high, which suggests a propensity to form initially liquid-like droplets that would subsequently age. Our findings also suggest that sequences with low complexity are prone to form ordered fibrils when exposed to environmental changes that increase the effective persistence length. The aging exponent ($\mu$), in this case, could exceed unity (see Fig.\ref{fig:Dynamics_epb=4}b), which has been observed in experiments~\cite{jawerth2020protein}. 

To synthesize our findings, we present a schematic diagram of states for protein aggregates in Fig.\ref{fig:State_Diagram}b. Because of the aging effects, we emphasize that it is not an equilibrium phase diagram, but rather is schematic reflecting behaviors that could occur on the experimental timescales. For highly flexible sequences with low complexity, aggregates exhibit a liquid-like ergodic behavior. As the effective rigidity is increased, these aggregates transition to ordered fibrils. However, for high-complexity sequences, this progression to ordered fibrils is hindered, primarily due to the dynamical arrest, leading to glassy dynamics, reflecting heterogeneous organization within the aggregates. Fig.\ref{fig:State_Diagram}b suggests that there is a spectrum of possibilities depending on the interplay between sequence complexity and monomer rigidity.  

\subsection*{Repeat RNA sequences} Experiment~\cite{jain2017rna} and simulations~\cite{nguyen2022condensates} have shown that RNA repeat sequences, (CAG)$_n$ and (CUG)$_n$ ($n$ is degree of polymerization) under {\it in vitro} conditions form condensates. The liquid condensates age, and reach a gel-like state over time~\cite{jain2017rna}.  Unlike the IDP sequences,  the ground states of (CAG)$_n$ monomer for a range of $n$ are near-perfect hairpins~\cite{maity2023odd}. More importantly, the tangent correlation function 
has the characteristics of semi-flexible polymers~\cite{nguyen2022condensates}. In other words, the repeat RNA low-complexity sequences are rigid, which in terms of the current model, implies that $\epsilon_b \ne 0$.  Because $H_3=1.58$ for (CAG)$_n$, and the monomer is rigid, we expect that they are likely to form ordered structures (Fig.\ref{fig:State_Diagram}b).  Our admittedly crude prediction is not inconsistent with {\it in vitro} experiments \cite{jain2017rna} showing solid-like material formation.   A  complication with RNA is that divalent cations (Mg$^{2+}$ for example) could decrease the persistence length, which could result in a transition from a partially ordered solid to an amorphous liquid-like structure. 

\subsection*{Conclusion}
We introduced a minimal model to glean insights into the interplay between sequence complexity and monomer rigidity on the morphology and dynamics of protein aggregates. By examining two extreme cases of low and high-complexity sequences and varying the persistence length of monomers, we discovered that morphologies and the aging dynamics of the aggregates are dramatically altered. We conclude with a few additional remarks. (1) The flexible low-complexity sequence forms liquid droplets whose relaxation times are independent of the waiting time, implying that their dynamics are ergodic.  In contrast, all other studied cases exhibit non-ergodic behavior, regardless of the sequence complexity or monomer bending rigidity. More importantly, these systems age, with the relaxation times that increase dramatically with the waiting times. (2) The low-complexity sequence, at non-zero values of the bending rigidities, forms ordered fibrils in which the monomers are arranged as in a nematic crystal. In contrast, the morphologies of the high-complexity sequence are amorphous, regardless of the value of the bending rigidity.  Although the dynamics of the ordered fibril and the disordered aggregates are qualitatively similar (they both display stretched exponential kinetics), the nature of ergodicity breaking in the two cases is different. In the fibril-forming low-complexity case, the dynamics of the aggregate are restricted to the basin corresponding to the low free energy fibril-like state, with a vanishingly low probability of escape. In contrast, when the sequence complexity is high, the system explores lower and lower free energy minima as the time waiting time grows. This corresponds to the  weak ergodicity breaking scenario discussed in the trap model~\cite{bouchaud1992weak}.  Our work shows that by exploring a wider range of parameter space (sequence complexity and bending energy) using minimal models, insights into the link between aging and morphology in protein condensates can be provided.


\matmethods{

\section*{Molecular Dynamics Simulation\label{sec:Energy Function}}
\textbf{Energy Function:}
The total potential energy of the polymer chain is, 
$U_{} =   \sum_{i<j} U_{\rm{NN}}(r_{i,j}) + \sum_{i} U_{\rm{FENE}}(r_{i,i+1}) +\sum_{k} U^{}_{\rm{ANG}}(\theta_{k}) $, where $i$ and $j$ are the indices of the beads, and $k$ is the angle between three successive beads in the monomer. 
We account for the sequence-dependent attraction between beads using  the potential,
\begin{equation} 
\label{eq:UNN}
U_{\rm{NN}}(r_{i,j})  = 4\epsilon_{} \Big( \big(\frac{\sigma}{r_{i,j}}\big)^{12}-\delta_{\alpha_i,\alpha_j} \big(\frac{\sigma}{r_{i,j}}\big)^{6} \Big) 
\Theta(r_c-r_{i,j}).
\end{equation}
For the ABC model, the interaction between the letters could be described using a 3$\times$3 matrix with six distinct interaction energy scales. For simplicity, we choose the interactions between all the beads to be the same. The parameter $\epsilon$, which is set to $\epsilon=2k_BT$ in Eq.\ref{eq:UNN}, is the energy scale in the Lennard-Jones potential, $\sigma$ is the range of the interaction that also sets the diameter of the beads in the simulations, $\alpha_i(\alpha_j) $ describes the type of bead $i(j)$ (A or B or C); $\delta_{\alpha_i,\alpha_j}$ is $1$ if bead $i$ is the same type as bead $j$, and 0 otherwise. Finally, $\Theta$ is the Heaviside step function used for the cutoff $r_c=4 \sigma$ for the potential. 

The second term in $U$ enforces the connectivity of the beads, and is given by,
\begin{align} 
\begin{split}
\label{}
U_{\rm{FENE}}(r_{i,i+1}) = -\frac{1}{2}k_F R_F^2 \log\Big[1-\frac{(r_{i,i+1}-r^0_{i,i+1})^2}{R_F^2}\Big],
\end{split}  
\end{align}
where $k_F$ is the stiffness of the potential, $R_F$ is the upper bound for the displacement, and $r^0_{i,i+1}$ is the equilibrium distance between the beads, $i$ and $i+1$, set as $r^0_{i,i+1}=\sigma$. We set $k_F=30\epsilon/\sigma^2$ and $R_F=1.5 \sigma$~\cite{midya2019phase}.

The last term in $U$ is the angle potential that controls the bending stiffness of the polymer. The potential $U^{}_{ANG}(\theta_{k})$ is taken as, 
\begin{equation} 
\label{}
U^{}_{\rm{ANG}}(\theta_{k})  =\epsilon_b(1+\cos \theta_k),
\end{equation}
where $\epsilon_b$, the energy scale for bending, is related to the persistence length of the homopolymer without attractive interaction as $l_p \simeq l_b\epsilon_b/k_BT$, where $l_b$ is the bond length of the beads in simulation~\cite{midya2019phase}. 

\textbf{Simulation Details:}
We employ low friction Langevin dynamics simulations using the OpenMM software package~\cite{eastman2017openmm}. The simulation time-step is  $\Delta t_L=0.01\tau_L$, with $\tau_L=\sqrt{m\sigma^2/k_BT}$. All the beads in the simulation are assigned a unit mass. The friction coefficient is chosen as $0.01/\tau_L$. Simulations are conducted in a cubic box with an edge length of $100 \sigma$ and periodic boundary conditions. 

In multi-chain simulations, $64$ identical monomers, each comprising of $84$ connected beads, are placed randomly within the box. Initially, configurations are generated by running multi-chain simulations without the sequence-dependent interaction term (setting $\delta_{\alpha_i,\alpha_j}=0$ in Eq.(\ref{eq:UNN})). 
Following this, simulations are executed with sequence-dependent interactions, starting from the randomized monomer configurations. Interactions between beads from different chains are identical to those within the same chain.

\textbf{Sequence generation:}
Starting with the low complexity sequence, (ABC)$_{28}$ (ABCABCABC...), we shuffled the sequence, accepting mutations that increase the $H_3$ value, defined in Eq.(\ref{eq:H}). We repeated the shuffling procedure until $H_3$ reached $4.75$, thus generating the high complexity sequences. An  example of such a sequence is,  
ABACBAACBBACABBCBBBAACCBBBCCABCCBACCAABABCABABCBCAA
CAAABBCBABBBAAAABCACACCCBCCCCACBC.

\textbf{End-to-end distance distribution:}
We use the mean-field expression for the end-to-end distance distribution, which is fairly accurate, for the semi-flexible polymer ~\cite{ha1995mean,thirumalai1997statistical} to extract the persistent length. In terms of the contour length of the polymer ($L$) and persistence length ($l_p$), the distribution reads,
\begin{align} 
\begin{split}
\label{eq:Pe}
P(R_{ee}) = \frac{4\pi N (R_{ee}/L)^2}{L(1-(R_{ee}/L)^2)^{9/2}} \exp\Big(-\frac{3t}{4(1-(R_{ee}/L)^2)}\Big),
\end{split} 
\end{align}
where $t=3L/2l_p$ and $N=\frac{4\alpha^{3/2}e^{\alpha}}{\pi^{3/2}(4+12\alpha^{-1}+15\alpha^{-2})}$ with $\alpha=3t/4$. 
The counter length of the monomer in the simulation is $84\sigma$. We fit the expression Eq.~(\ref{eq:Pe}) to estimate the single parameter, $l_p$.


\section*{Data Analysis}
\textbf{Nematic order parameter $\bm{(S)}$:}
We calculated the nematic order parameter using the eigenvector associated with the smallest eigenvalue of the inertia tensor, 
\begin{equation} 
\label{}
I_{\alpha,\beta}=\sum_{i=1}^{L}m_i(r_i^2\delta_{\alpha,\beta}-r_{i,\alpha}r_{i,\beta}),
\end{equation}
defining the molecular axis of the monomer; $\mathbf{r}$ and $m$ are the positions of the beads relative to the center-of-mass of the monomer and the masses of the atoms, respectively. $\alpha$ and $\beta$, which take on values $1,2,3$, are the spacial indices. $L$ is the number of beads in a monomer. 
We define $S$ as the largest eigenvalue of the second rank  tensor,
\begin{equation} 
\label{}
Q_{\alpha,\beta}=\frac{1}{2M}\sum_{i=1}^{M}(3\hat{u}_{i,\alpha}\hat{u}_{i,\beta}-\delta_{\alpha,\beta}),
\end{equation}
where $\hat{u}$ is the unit vector of the molecular axis, computed from $I_{\alpha,\beta}$, and $M$ is the number of monomers in the simulation. We computed the nematic order parameter~\cite{allen2017computer} using python library MDTraj~\cite{mcgibbon2015mdtraj}.\\
\textbf{Radial distribution function ($\bm{g(r)}$):}
To compute $g(r)$, we first enumerated the radial pair-distances between the beads, $r$. We obtained the probability density of $r$ in the aggregate denoted as $n(r)$.  
Radial distribution function can be computed as 
\begin{equation} 
\label{}
g(r)=n(r)/(4\pi r^2 dr \rho),
\end{equation}
where $\rho=\sum_{r}n(r)/\frac{4\pi r^3}{3}$ is the local density. \\
\textbf{Radius of gyration ($\bm{R_G}$):}
The radius of gyration of the aggregates is computed using, 
\begin{equation} 
\label{}
R_G^2=\frac{1}{N}\sum_{i=1}^{N}(\bm{r}_i-\bm{r}_{cm})^2,
\end{equation}
where $\bm{r}_i$ is the position of $i^{th}$ bead, $\bm{r}_{cm}$ is the center of mass of the aggregate, and $N=L\times M$ is the total number of beads.  
We use MDTraj~\cite{mcgibbon2015mdtraj} to compute the radius of gyration. 

}

\showmatmethods{} 

\acknow{We are grateful to Balaka Mondal and Farkhad Maksudov for useful comments on the manuscript. This work was supported by a grant from the National Science Foundation (CHE 2320256) and the Welch Foundation through the Collie-Welch Chair (F-0019).}

\showacknow{} 

\bibsplit[2]


\bibliography{mybib.bib}

\end{document}


\pagebreak
\widetext
\begin{center}
\textbf{\large Supplemental Materials: Sequence Complexity and Monomer Rigidity Control the Morphologies and Aging Dynamics of Protein Aggregates}
\end{center}

\author{Ryota Takaki}
\affiliation{%
 Max Planck Institute for the Physics of Complex Systems 
}%


\author{D. Thirumalai}%
\email{dave.thirumalai@gmail.com}
\affiliation{%
 Department of Chemistry, The University of Texas at Austin, Austin, TX,78712
}%
\affiliation{Department of Physics, The University of Texas at Austin, Austin, TX,78712}

\setcounter{equation}{0}
\setcounter{figure}{0}
\setcounter{table}{0}
\setcounter{page}{1}
\makeatletter
\renewcommand{\theequation}{S\arabic{equation}}
\renewcommand{\thefigure}{S\arabic{figure}}
\renewcommand{\bibnumfmt}[1]{[S#1]}
\renewcommand{\citenumfont}[1]{S#1}
\maketitle
\section{Dynamics of Morphological Transitions}
\begin{figure}[h]
\centering
\includegraphics[width=0.8\textwidth]{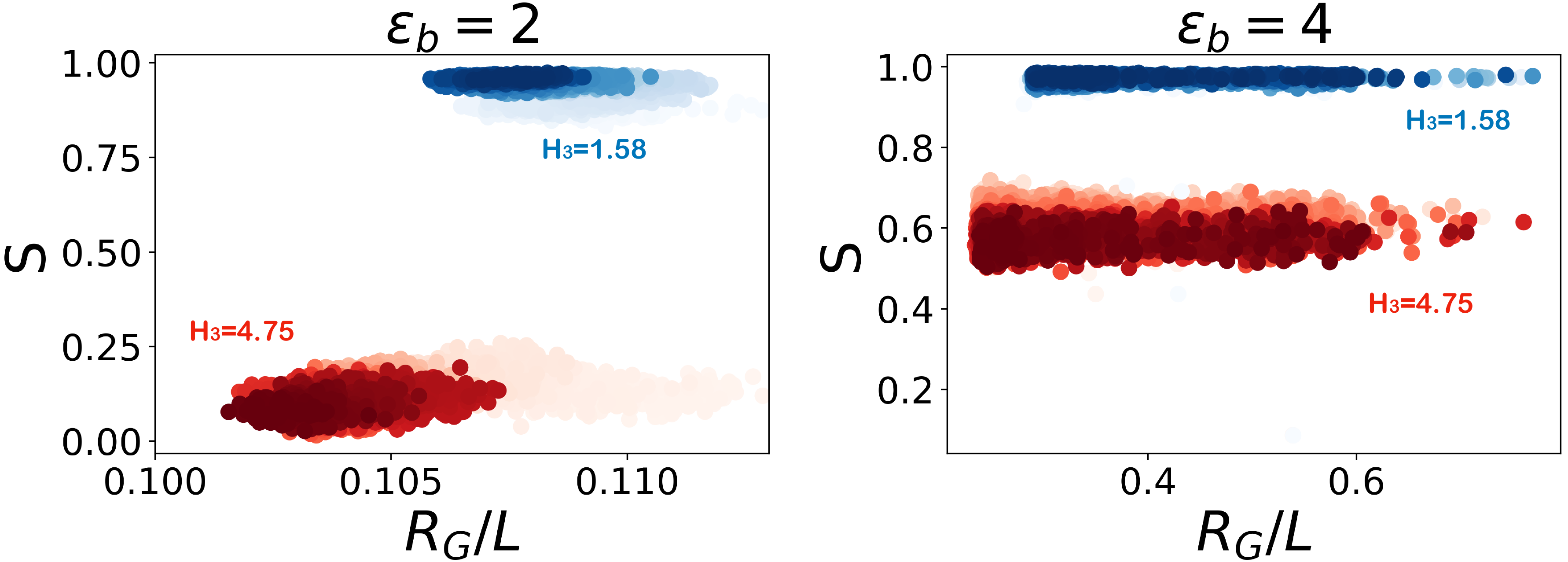}
\caption{\label{fig:phase_high_ep}{\bf Morphological transitions as a function of time for $\bm{\epsilon_b=2}$ and $\bm{\epsilon_b=4}$:} Trajectories illustrate the evolution of the morphology of aggregates. The $x$-axis represents the radius of gyration of aggregates, $R_G$, normalized by the length of the periodic box, $L=100\sigma$. The $y$-axis shows the nematic order parameter, $S$. The opacity of the dots indicates progression in time, with denser dots corresponding to more recent times. Blue and red dots correspond to $H_3=1.58$ and $H_3=4.75$, respectively. }
\end{figure}

We show the morphological transition of aggregates in time at in stiff chains for $H_3=1.58$ and $H_3=4.75$ in Fig.\ref{fig:phase_high_ep}. The corresponding results for $\epsilon_b=0$ are shown in Fig. 2d in the main text.   Higher nematic order parameters for $H_3=1.58$ illustrate the fibril formation. In contrast,  the $H_3=4.75$ sequence at $\bm{\epsilon_b=2}$ is disordered with low nematic order. At $\bm{\epsilon_b=4}$, the value of the nematic order parameter is non-zero (left panel in Fig.\ref{fig:phase_high_ep}).  The conformational fluctuations are greater in the high-complexity sequence compared to the low-complexity sequence.   There is signature of aging for both the high and low complexity sequences for the higher rigidities. 

\section{pair Mean Squared Displacement for higher rigidities}
\begin{figure}[h]
\centering
\includegraphics[width=0.7\textwidth]{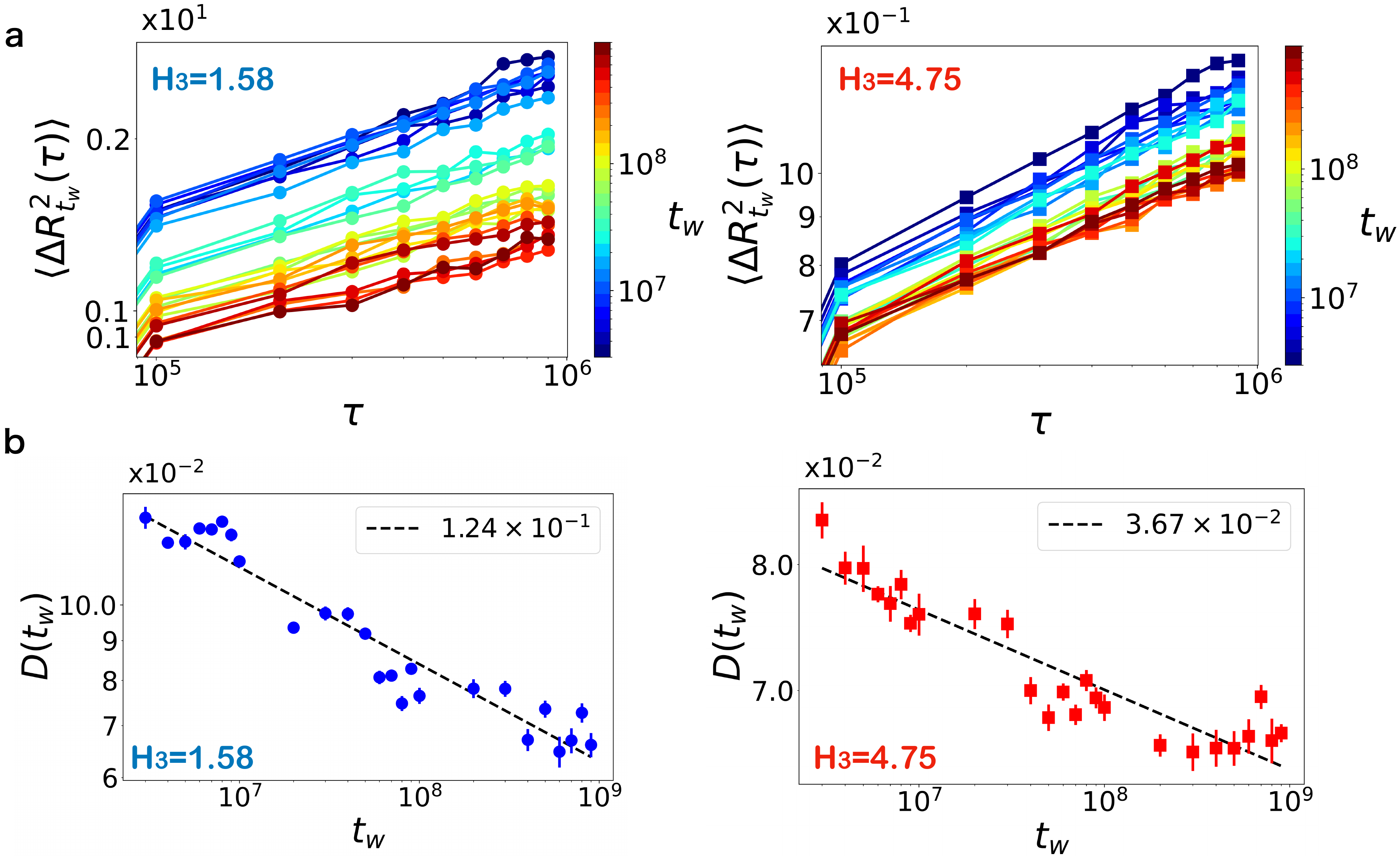}
\caption{\label{fig:pMSD_epb=2}{\bf pMSD for $\bm{\epsilon_b=2}$:} (a) pMSD  for low complexity sequence ($H_3=1.58$, left) and high complexity sequence ($H_3=4.75$, right). The color gradient indicates the system's age ($t_w$). In contrast to Fig.3 in the main text, pMSD for both $H_3$ values depend on the waiting time, $t_w$.
(b) The generalized diffusion coefficient $D(t_w)$ as a function of the age $t_w$ obtained by fitting Eq.(5) in the main text to Fig.\ref{fig:pMSD_epb=2}(a).  The error bars for $D(t_w)$ are the standard errors. The dashed line is a power-law fit with exponent $\alpha=(1.24\pm 0.16)\times 10^{-1}$ for $H_3=1.58$ and $\alpha=(3.67\pm 0.66)\times 10^{-2}$ for $H_3=4.75$. The error in the exponents is the $95 \%$ confidence interval.
The ensemble average is performed over $10$ shifted time windows between the successive $t_w$s and $2$ different trajectories, generating $20$ ensembles.  }
\end{figure}

\begin{figure}[h]
\centering
\includegraphics[width=0.7\textwidth]{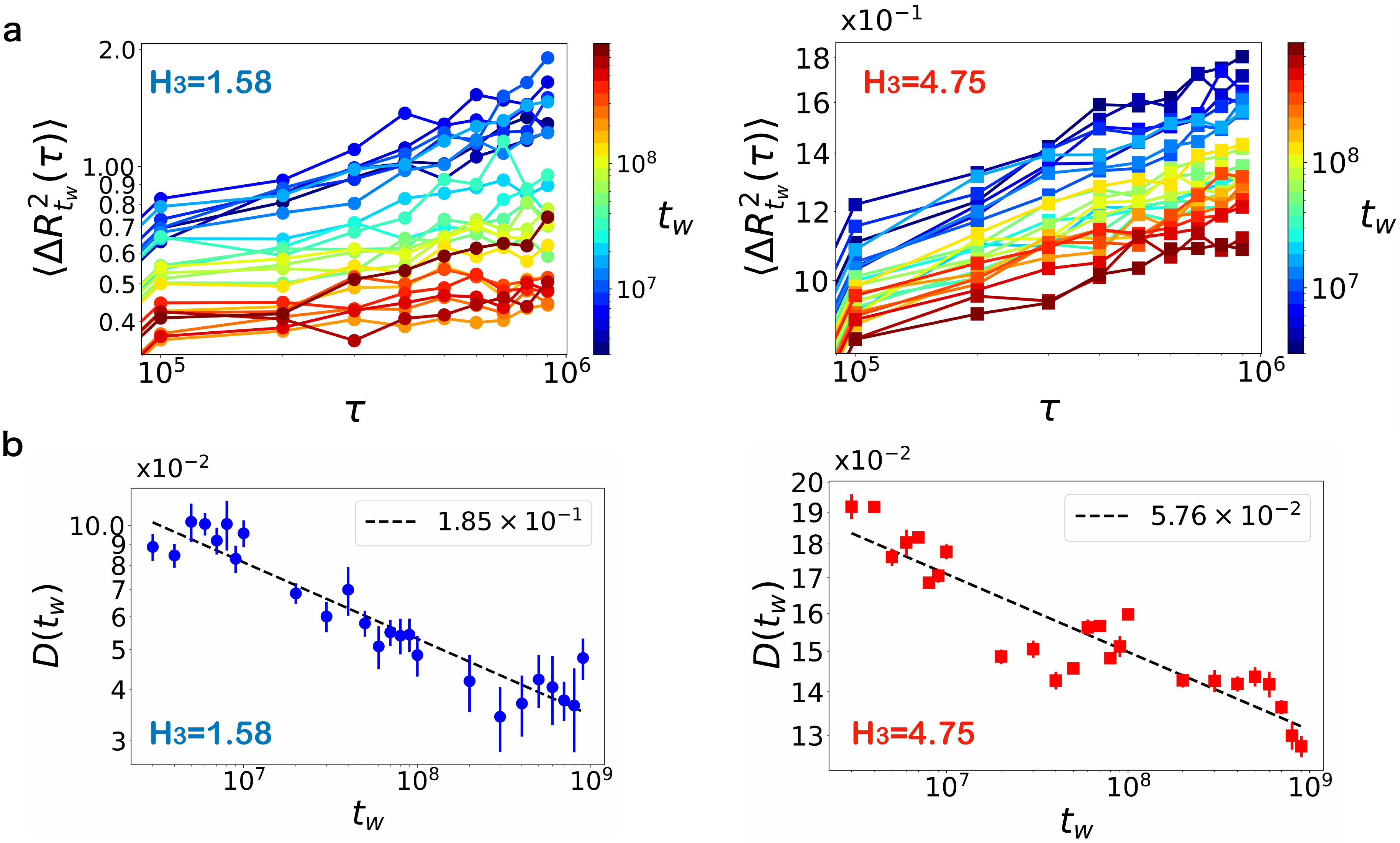}
\caption{\label{fig:pMSD_epb=4}{\bf pMSD for $\bm{\epsilon_b=4}$:} (a) pMSD  for low complexity sequence ($H_3=1.58$, left) and high complexity sequence ($H_3=4.75$, right). The color gradient indicates the system's age ($t_w$). 
(b) Same as Fig. \ref{fig:pMSD_epb=2}, except the results are for $\epsilon_b=4$. 
The dashed line is a power-law fit with exponent $\alpha=(1.85 \pm 0.28) \times 10^{-1}$ for $H_3=1.58$ and $\alpha=(5.76 \pm 1.09)\times 10^{-2}$ for $H_3=4.75$.} 
\end{figure}

We calculated the pMSD for higher rigidities: $\epsilon_b=2$ and $\epsilon_b=4$. In accord with the results in the main text (Fig.5-6), we find evidence for aging dynamics in pMSD for higher rigidities (Fig.\ref{fig:pMSD_epb=2}-\ref{fig:pMSD_epb=4}).  
\section{Connection to the trap model}
\begin{figure}[h]
\centering
\includegraphics[width=0.7\textwidth]{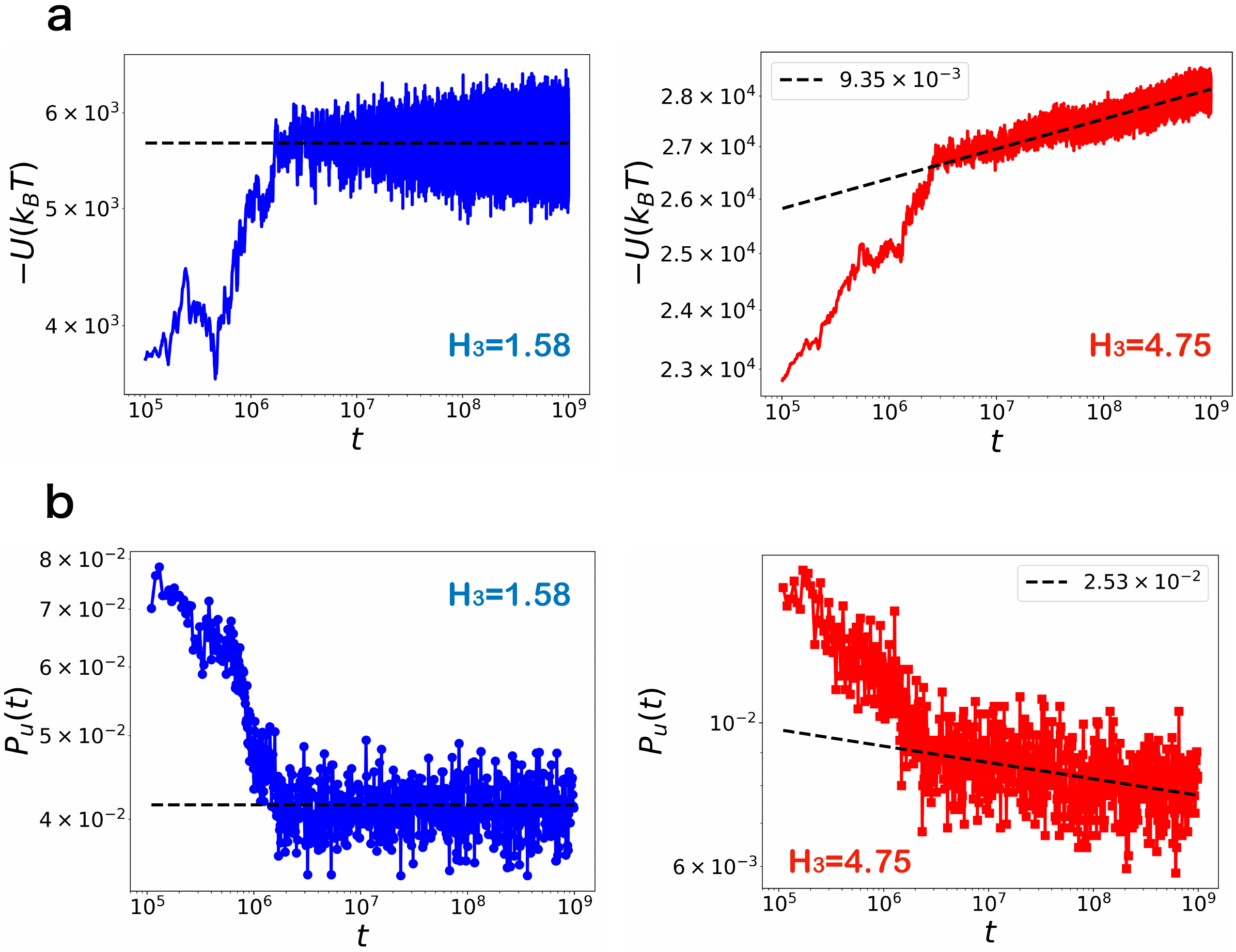}
\caption{\label{fig:TrapModel_Connection}{\bf Glassy behavior in the simulation is consistent with the trap model:} (a) Potential energy ($-U$) in the simulation trajectory as a function of time for $\epsilon_b=0$. The left and right figure show $H_3=1.58$ and $H_3=4.75$, respectively. The black dashed line is the fit for $t>10^7$. The line is a constant for $H_3=1.58$ and the power-law with the exponent $(9.35 \pm 0.81)\times 10^{-3}$ for $H_3=4.75$. (b) The unbound probability $P_u(t)$ for $\epsilon_b=0$. The left and right figure show $H_3=1.58$ and $H_3=4.75$, respectively. The line is a constant for $H_3=1.58$ and the power-law with the exponent $-(2.53 \pm 0.81) \times 10^{-2}$ for $H_3=4.75$. The errors of the exponents are the $95 \%$ confidence interval of the estimates. 
}
\end{figure}

The theory of aging protein in condensates~\cite{takaki2023theory} is based on the trap model, which first proposed to describe a similar phenomenon in spin glasses~\cite{bouchaud1992weak,monthus1996models}. In essence, aging is attributed to the diminishing fraction of diffusive elements, $P_u(t)$, as the condensate ages, leading to slow dynamics. Although the increased rigidity of monomers adds to the complexity of aging protein condensates, our simulations show that key aspects of the trap model hold. The trap model predicts that, during aging, the system is progressively trapped in deeper energy wells over time. This is illustrated in Fig.~\ref{fig:TrapModel_Connection}a, which plots the potential energy in the simulation trajectory ($U$; see Methods section in the main text). The low complexity sequence exhibits equilibrium behavior, in contrast to the power-law increase  of $-U$ observed for the high complexity sequence. Similar behavior has been found in simulations of simple glass-forming systems~\cite{kob1997aging,Angelani01PRL}. 

We also calculate the probability $P_u(t)$, representing the fraction of beads within aggregates not connected to any others (Fig.~\ref{fig:TrapModel_Connection}b with $\epsilon_b=0$). The connection cut-off is set at $1.5 \sigma$, which excludes trivial adjacent bonds in the monomers. For $H_3=1.58$ sequence, $P_u(t)$ demonstrates equilibration, whereas for the $H_3=4.75$ sequence, a power-law decay is observed as anticipated by the theoretical study~\cite{takaki2023theory}. 

\section{Videos}
Supplementary movie 1: Initial stage ($t \leq 3 \times 10^6$ steps) of the ordered aggregate formation for the parameters $H_3=1.58$ and $\epsilon_b=2$. Bead colors denote the sequence letters, with red corresponding to A, green to B, and blue to C. Initially liquid-like and disordered aggregate forms. Subsequently, an ordered structure starts to appear.

Supplementary movie 2: Initial stage ($t \leq 3 \times 10^6$ steps) of the ordered aggregate formation for the parameter $H_3=1.58$ and $\epsilon_b=4$. Bead colors denote the sequence letters, with red corresponding to A, green to B, and blue to C. Ordered and elongated structure appears rapidly after  aggregation. 

\bibliography{mybib}